\documentclass{article}

\usepackage{arxiv}
\usepackage{enumitem}
\usepackage[utf8]{inputenc} % allow utf-8 input
\usepackage[T1]{fontenc}    % use 8-bit T1 fonts
\usepackage{hyperref}       % hyperlinks
\usepackage{url}  
\usepackage{natbib}% simple URL typesetting
\usepackage{booktabs}       % professional-quality tables
\usepackage{amsfonts}       % blackboard math symbols
\usepackage{nicefrac}       % compact symbols for 1/2, etc.
\usepackage{microtype}      % microtypography
\usepackage{lipsum}
\usepackage{graphicx}
\graphicspath{ {./images/} }
\usepackage{pdfpages}
\usepackage{amsmath}
\usepackage{rotating}
\usepackage{epstopdf}
\usepackage{lscape}
\usepackage{alltt}
\usepackage{booktabs, subfigure}
\usepackage{verbatim, setspace}
\usepackage{calc, color, setspace}
\usepackage[english]{babel}
\usepackage{tabularx}
\usepackage{longtable}
\usepackage{multirow}
\usepackage{booktabs}
\usepackage{ragged2e}
\usepackage{subfigure}
\usepackage{tikz}
\usepackage[none]{hyphenat}
\usepackage{verbatim}
\usepackage{ragged2e}
\usepackage{breakcites}

\usepackage{times}

\newcommand{\bmu}{\mbox{\boldmath $\mu$}}

\newcommand{\bSigma}{\mbox{\boldmath $\Sigma$}}
\newcommand{\bepsilon}{\mbox{\boldmath $\epsilon$}}

\newcommand{\bbeta}{\mbox{\boldmath $\beta$}}
\newcommand{\balpha}{\mbox{\boldmath $\alpha$}}
\newcommand{\btheta}{\mbox{\boldmath $\theta$}}

\newcommand{\bV}{\mathbf{V}}
\newcommand{\bPsi}{\mbox{\boldmath $\Psi$}}

\newcommand{\bOmega}{\mbox{\boldmath $\Omega$}}
\newcommand{\bomega}{\mbox{\boldmath $\omega$}}

\newcommand{\bC}{\mathbf{C}}

\newcommand{\z}{\mathbf{Z}}

\newcommand{\Z}{\mathbf{Z}}
\newcommand{\R}{\mathbf{R}}
\newcommand{\X}{\mathbf{X}}

\title{Spatial Censored Regression Models in R: The {{CensSpatial}}\hspace{0.1cm} Package}

\author{
Jos\'e Alejandro Ordo\~nez \\
  Department of Statistics\\
  State University of Campinas\\
  Campinas, Brazil 13083-970 \\
  \texttt{ordonezjosealejandro@gmail.com} \\
   \And
  Christian E. Galarza \\
  Department of Mathematics\\
  Escuela Superior Polit\'ecnica del Litoral\\
  Guayaquil, Ecuador 090112 \\
  \texttt{chedgala@espol.edu.ec} \\
  \And
 Victor H. Lachos \\
  Department of Statistics \\
  University of Connecticut\\
  Connecticut, USA 06269  \\
  \texttt{hlachos@uconn.edu} \\
}

\begin{document}

\maketitle
\begin{abstract}
{\texttt{CensSpatial} is an R package for analyzing spatial censored data through linear models. It offers a set of tools for simulating, estimating, making predictions, and performing local influence diagnostics for outlier detection. The package provides four algorithms for estimation and prediction. One of them is based on the stochastic approximation of the EM (SAEM) algorithm, which allows easy and fast estimation of the parameters of linear spatial models when censoring is present. The package provides worthy measures to perform diagnostic analysis using the Hessian matrix of the completed log-likelihood function. This work is divided into two parts. The first part discusses and illustrates the utilities that the package offers for estimating and predicting spatial censored data. The second one describes the valuable tools to perform diagnostic analysis. Several examples in spatial environmental data are also provided.} 
\end{abstract}

% keywords can be removed

\section{Introduction}

Spatial data can be described as observations that are spatially located in a coordinated area. Their analysis constitutes an area of statistics that is characterized by the spatial correlation among individuals where the interest is to predict the response at non-sampling sites. Usually, spatial analysis assumes that observations are fully observed, but this is not always possible. Sometimes data are subject to  upper or lower (or both) detection limits, beyond which they are not quantifiable \citep{Schelin}. An example of this can be seen in environmental (spatial) monitoring of different variables which often involves left-censored observations falling below the minimum limit of detection (LOD) of
the instruments used to quantify them.

Methods for handling this type of data have been widely discussed in the statistical literature. \cite{rathbun2006spatial}
applied a Robbins-Monro stochastic approximation to estimate the parameters of a spatial regression model, obtain conditional simulations of left-censored observations and then obtain a predictor for data at unsampled sites by taking the weighted mean of kriging predictors via importance sampling. \cite{Schelin} proposed a semi-naive method that determines imputed values at censored locations in an iterative algorithm together with variogram estimation. They compared its predictive performance through a simulation study with Rathbun's, Naive 1 and Naive 2 methods, which involve replacing the censored data by the LOD or the LOD/2 respectively. \cite{militino1999analyzing} developed an EM-type algorithm for maximum likelihood (ML) estimation in censored spatial
data. However, this approach suffers from several drawbacks that
restrict its applicability, such as the computation of non-closed form expressions, which depend on high dimensional integrals.

%For instance, \cite{de2005bayesian}
%noted that this ML approach does not provide a means to estimate the
%correlation structure in the data and hence assumes it is known.\\

The EM algorithm proposed by \cite{Dempster77} is an iterative procedure for finding the maximum likelihood (ML) estimators of models that involves unobserved latent variables. In our context, an EM-type algorithm uses the observed and censored data to obtain estimates from the mean and correlation structures of a spatial model. In some cases, this algorithm cannot be carried out because the calculation of the expectations involved in E-step cannot be computed analytically. An alternative is to approximate those expectations by Monte Carlo methods to replace the traditional E-step. This procedure is the so-called Monte Carlo Markov chain (MCEM) algorithm. Even though it is possible to approximate these expectations, this can involve heavy computational effort when some generating methods are used to sample from the conditional latent variable distributions at each iteration.

An alternative to deal with this problem is the SAEM algorithm proposed by \cite{delyon1999convergence}. This algorithm replaces the E-step by a stochastic approximation obtained using simulated data, while the M-step remains unchanged. In the framework of spatial models, \cite{jank2006implementing} showed that the computational effort of
SAEM is much smaller and it reaches convergence in just a fraction of
the simulation iterations when compared with the MCEM algorithm. This is due to the fact that a memory effect is contained in the SAEM method, in which the
previous simulations are considered in the computation of the posterior ones.

Another important topic in spatial censored data analysis is the study of influential observations. Influence analysis is an important statistical tool because it can
assess the goodness-of-fit and influential observations that can distort the parameter estimates leading in some cases to erroneous inference \citep{gimenez}. For influential observation detection, we discuss two approaches; the first one is the case-deletion approach \citep{cook1977detection}, an intuitively appealing method which has been applied to many statistical models. The second one is local influence diagnostics \citep{cook86}. This technique assesses the stability of the estimation outputs with respect to the model inputs and has recently received attention in the literature on spatial models \citep[see, e.g.,][]{de2014influence}.  \cite{ZhuLee2001} developed an approach to perform local influence analysis for general statistical models with missing data, based on the {\it Q}-displacement function which is closely related to the conditional expectation of the complete-data log-likelihood in the E-step of the EM algorithm. This approach produces results very similar  to those obtained from  Cook's method. Moreover, the case-deletion  can be studied by the {\it Q}-displacement function, following the approach of \cite{zhu2001case}. So, we develop here methods to obtain case-deletion measures and local influence measures by using  the method of \cite{zhu2001case} \citep[see also][]{LeeXu04,ZhuLee2001} in the context of spatial censored models.
		
The aim of this paper is to present the utilities of the R package \texttt{CensSpatial}. The rest of the paper is organized as follows. In Section \ref{s1}, we derive all the theoretical framework for the estimation and prediction functions for spatial censored data implemented in the proposed package, along with the R functions for calling these methods. Also we present some useful graphical tools for interpretation as well as an application to real data. Section \ref{s2} presents the local influence diagnostic methods to detect influential observations. Graphical tools and an application are also provided in this section. Further details such as first and second derivatives of the spatial correlation, can be found in the Appendix.

\section{Estimation and prediction utilities}\label{s1}

First, we specify the spatial censored linear (SCL) model in a general sense and then we describe all the estimation and prediction tools implemented in the \texttt{CensSpatial} package.% We use the Rainfall average %dataset, available in the \texttt{CensSpatial} package, to illustrate our  proposed methods.

%%%%%%%%%%%%%%%%%%%%%%%%%%%%%%%%%%%%%%%%%%%%%%%%%%%%%%%%%%%%%%%%%%%%%%%%%%%%%%%
%%%%%%%%%%%%%%%%%%%%%%%%%%%%%%%%%%%%%%%%%%%%%%%%%%%%%%%%%%%%%%%%%%%%%%%%%%%%%%%
\subsection{Background of spatial linear models for censored response}

As in \cite{de2014influence}, we will consider a Gaussian model with a linear specification for
the spatial trend, which allows the inclusion of polynomial trends or more generally, spatially referenced covariates. The linear model is defined as

\begin{equation}\label{linearSPM}
\Z=\X\bbeta+\bepsilon,
\end{equation}

where $\X$ and $\bbeta$ represent the trend matrix and the parameters considered for the mean structure respectively, and $\bepsilon\sim N_n(\mathbf{0},\bSigma)$ is the stochastic component. We consider for the spatial case, $\X$ is a full rank matrix and  $\bSigma$ is a non-singular matrix with the form  $\bSigma=\sigma^2\bPsi$, with $\bPsi=(\nu^2\mathbf{I}_n+\R(\phi))$, where $\mathbf{I}_n$ represents an $n \times n$ identity matrix,
$\R(\phi)$ is the correlation matrix for the error term and $\nu^2=\tau^2/\sigma^2$ (see Appendix). We also assume that the response $Z_{i}$ is not fully observed for all area $i$, so we call $(V_{i},C_i)$ the observed data for the $i$th area, while $V_i$ represents either an uncensored observation $(V_i=V_{0i})$ or the LOD of censoring level $(V_i=(V_{1i},V_{2i}))$ and $C_i$ is the censoring indicator such that:
\begin{eqnarray}\label{linearSPM2}
C_i =
\left\{\begin{array}{ccc}
  1 & if & V_{1i}\leq Z_{i}\leq V_{2i}\,, \\
  0 & if & Z_{i} = V_{0i}\,.
  \end{array}\right.
\end{eqnarray}
Note that when $C_i=1$, if $Z_i \in (-\infty,V_{2i}]$, then we get a left censored SCL model \citep{toscas2010spatial}, and if $Z_i \in [V_{1i},\infty)$, then we get a right censored SCL model. The model defined in (\ref{linearSPM})-(\ref{linearSPM2}) is called the spatial censored
linear (SCL) model.

\section{Likelihood and SAEM implementation}\label{sec:loglik}

To compute the likelihood function associated with the SCL model, the observed and censored components of $\mathbf{Z}$ must be treated separately.  Let $\Z^o$ be the $n^o$-vector of observed responses and $\Z^c$ be
the $n^c$-vector of censored observations, with $(n=n^o+n^c)$, such that $C_{i}=0$ for all elements in $\Z^o$, and $C_{i}=1$ for all elements in $\Z^c$. After reordering, $\Z$, $\textbf{V}$, $\X$, and $\bSigma$ can be partitioned as follows:

$$\Z=vec(\Z^o,\Z^c), \quad \textbf{V}=vec(\bV^o,\bV^c),\quad
\X^{\top}=[\X^{o\top},\X^{c\top}] \quad  {\rm and}\quad
\bSigma=\left[\begin{array}{cc}
       \bSigma^{oo} & \bSigma^{oc} \\
       \bSigma^{co} & \bSigma^{cc}
     \end{array}\right],$$
where $vec(\cdot)$ denotes the function which stacks vectors or matrices having the same number of columns. Consequently, $\Z^o\sim N_{n^o} (\X^o\bbeta,\bSigma^{oo})$,{{ $\Z^c|\Z^o\sim N_{n^c}(\bmu,\mathbf{S}),$}} where $\bmu=\X^c\bbeta+\bSigma^{co}(\bSigma^{oo})^{-1}(\Z^o-\X^o\bbeta)$
and $\mathbf{S}=\bSigma^{cc}-\bSigma^{co}(\bSigma^{oo})^{-1}\bSigma^{oc}$. Now, let $\phi_{n}(\mathbf{u};\mathbf{a},\mathbf{A})$ be the \emph{pdf} of $N_{n}(\mathbf{a},\mathbf{A})$ evaluated at $\mathbf{u}$. From \cite{vaida2009fast} and \cite{jacqmin2000analysis}, the likelihood function (using conditional probability arguments) is given by
\begin{eqnarray}
L(\btheta)=\phi_{n^o}(\mathbf{Z}^o;\X^o\bbeta,\bSigma^{oo})P(\mathbf{Z}^c \in \mathbb{V}^c |\mathbf{Z}^o),
\end{eqnarray}
where
$$\mathbb{V}^c=\{\Z^c=(Z_{1}^c,\ldots,Z_{n^c}^c)^{\top}|V_{11}^c\leq
Z_{1}^c\leq V_{21}^c,\ldots,V_{1n^c}^c\leq Z_{n^c}^c\leq
V_{2n^c}^c\}$$
and $P(\mathbf{u} \in \mathbb{A}|\Z^o)$ denotes the conditional probability of $\mathbf{u}$ being in the set $\mathbb{A}$ given the observed response. This function can be evaluated without much computational burden through the {\texttt{mvtnorm}} routine available in the namesake R package \citep[see][]{genz2008mvtnorm}. Now, we can compare models using a likelihood based criterion, such as the Akaike \citep[AIC;][]{Akaike74} and Schwarz \citep[BIC;][]{Schwarz1978} information criteria.

\subsection{SAEM algorithm for censored spatial data}\label{SAEMspatial}
We propose an EM-type (SAEM) algorithm by considering $\Z$
as a missing data or latent variable. In the estimation, the reading $\bV$ at the censored observations  ($C_i=1$) is treated as hypothetical missing data, and augmented with the observed dataset $\Z_c=(\bC^{\top},\bV^{\top},\Z^{\top})^{\top}$. Hence, the
complete-data log-likelihood function is given by:
\begin{eqnarray}
\ell_c(\btheta)&\propto&-\frac{1}{2}\left[\log(|\bSigma|)+(\Z-\X\bbeta)^{\top}\bSigma^{-1}(\Z-\X\bbeta)\right]+cte,
\end{eqnarray}

with $cte$ being a constant, independent of the parameter vector \btheta. Given the current estimate $\btheta=\widehat{\btheta}^{(k)}$, the E-step computes the
conditional expectation of the complete data log-likelihood function, i.e., $Q(\btheta|\widehat{\btheta}^{(k)}) = E[\ell_c(\btheta|\mathbf{Z}_c)|\bV,\bC,\widehat{\btheta}^{(k)}]$. Denote the two conditional first moments for the response $\Z$ as $\widehat{\Z}^{(k)}=E\{\displaystyle\Z|\textbf{V},\bC,\widehat{\btheta}^{(k)}\}$ and $\widehat{\Z\Z^{\top}}^{(k)}=E\{\displaystyle\Z\Z^{\top}|\textbf{V},\bC,\widehat{\btheta}^{(k)}\}$. For the SCL model, we have

\begin{eqnarray*}
Q(\btheta|\widehat{\btheta}^{(k)})&=& -\frac{1}{2}\left[\log(|\bSigma|)+ \widehat{A}^{(k)}\right],
\end{eqnarray*}
where
$$\widehat{A}^{(k)}=
tr\big(\widehat{\Z\Z^{\top}}^{(k)}\bSigma^{-1}\big)-2\widehat{\Z}^{(k)\top}\bSigma^{-1}\X\bbeta+\bbeta^{\top}\X^{\top}\bSigma^{-1}\X\bbeta.$$

\begin{comment}
It is clear that the E-step reduces only to the computation of
\begin{equation}\label{MomenNormal}
\widehat{\Z}^{(k)}=E\{\displaystyle\Z|\textbf{V},\bC,\widehat{\btheta}^{(k)}\}
\mbox{\quad and\quad}
\widehat{\Z\Z^{\top}}^{(k)}=E\{\displaystyle\Z\Z^{\top}|\textbf{V},\bC,\widehat{\btheta}^{(k)}\}.
\end{equation}
\end{comment}

\noindent $\Z$ is not fully observed, so the components of $\widehat{\Z}^{(k)}$ and $\widehat{\Z\Z}^{(k)}$, corresponding to $C_i=1$ will be estimated by the first two moments of the truncated normal distribution respectively. When $C_i=0$, these components can be obtained directly from the observed values, i.e, $\widehat{\Z}^{(k)}=\Z^o$ and $\widehat{\Z\Z}^{(k)}=\Z^o\Z^{o\top}$. Although these expectations exhibit closed forms (as functions of multinormal  probabilities; for further details, please refer to \cite{arismendi}), the calculation is computationally expensive requiring high-dimensional numerical integrations, resulting in convergence issues when the proportion of censored observations is non-negligible. At each iteration, the SAEM algorithm successively simulates from the conditional distribution of the latent variable, and updates the unknown parameters of the model. Thus, at iteration $k$ , the SAEM proceeds as follows:

\begin{itemize}
\item \textbf{Step E-1 (sampling)}: Sample $\Z^{c}$ from a truncated normal distribution, denoted by $TN_{n^{c}}(\bmu,\mathbf{S};\mathbb{A}^c)$, with
$\mathbb{A}^c=\{\Z^c=(Z_{1}^c,\ldots,Z_{n^c}^c)^{\top}|V_{11}^c\leq
Z_{1}^c\leq V_{21}^c,\ldots,V_{1n^c}^c\leq Z_{n^c}^c\leq
V_{2n^c}^c\},$
$\bmu=\X^c\bbeta+\bSigma^{co}(\bSigma^{oo})^{-1}(\Z^o-\X^o\bbeta)$
and
$\mathbf{S}=\bSigma^{cc}-\bSigma^{co}(\bSigma^{oo})^{-1}\bSigma^{oc}$.
Here $TN_{n}(.;\mathbb{A})$ denotes the $n$-variate truncated normal distribution in the interval $\mathbb{A}$, where $\mathbb{A}=A_{1}\times\ldots\times A_{n}$. The new observation $\Z^{(k,l)} = (Z^{(k,l)}_{1},\ldots,Z^{(k,l)}_{n^{c}},Z_{n^{c}_i+1},\ldots,Z_{n})$ is a sample generated for the $n^{c}$ censored cases and the observed values (uncensored cases), for $l=1,\ldots, M.$

\item \textbf{Step E-2 (stochastic approximation)}: Since we have the sequence $\Z^{(k,l)}$, at the $k$-th iteration we replace the conditional expectations $\widehat{\Z}^{(k)}$ and $\widehat{\Z\Z^{\top}}^{(k)}$ by the following stochastic approximations:
\end{itemize}

\begin{eqnarray}\label{eq_saem}
\widehat{\Z}^{(k)} & = & \widehat{\Z}^{(k-1)} + \delta_k
\left[\frac{1}{M}\sum_{l=1}^M \Z^{(k,l)}
-\widehat{\Z}^{(k-1)}\right],\\
\widehat{\Z\Z^{\top}}^{(k)} & = & \widehat{\Z\Z^{\top}}^{(k-1)} +
\delta_k \left[\frac{1}{M}\sum_{l=1}^M \Z^{(k,l)}\Z^{(k,l)\top} -\widehat{\Z\Z^{\top}}^{(k-1)}\right],
\end{eqnarray}

where $\delta_{k}$ is a smoothness parameter, i.e., a decreasing sequence of positive numbers as defined in \cite{kuhn2004coupling}, such that
$\sum_{k=1}^{\infty}\delta_{k}=\infty$ and $\sum_{k=1}^{\infty}\delta_{k}^{2}<\infty$. For the SAEM, the E-Step coincides with the MCEM algorithm at the expense of a significantly smaller number of simulations $M$ (suggested to be $M\leq20$). For more details, see \cite{delyon1999convergence}.\\

\noindent Finally, the conditional maximization (CM) step maximizes $Q(\btheta|\widehat{\btheta}^{(k)})$ with
respect to $\btheta$ and obtains a new estimate $\widehat{\btheta}^{(k+1)}$, as follows:\\

\textbf{CM-Step (conditional maximization)}:

\begin{eqnarray}
\widehat{\bbeta}^{(k+1)}&= & \big(
\X^{\top}\widehat{\bSigma}^{-1(k)}\X\big)^{-1}
\X^{\top}\widehat{\bSigma}^{-1(k)}\widehat{\Z}^{(k)},\nonumber\\
\widehat{\sigma^2}^{(k+1)}&=&  \frac{1}{n}\Big[tr\big(\widehat{\Z\Z^{\top}}^{(k)}\widehat{\bPsi}^{-1(k)}\big)-2\widehat{\Z^{\top}}^{(k)}\widehat{\Psi}^{-1(k)}\X\widehat{\bbeta}^{(k+1)}+\widehat{\bbeta}^{\top(k+1)}\X^{\top}\widehat{\bPsi}^{-1(k)}\X\widehat{\bbeta}^{(k+1)}\Big]\,\,\nonumber\\
\widehat{\balpha}^{(k+1)}&=& \underset{\scriptscriptstyle \balpha \in
\mathbb{R}^+\times
\mathbb{R}^+}{\mathrm{argmax}}\left(-\frac{1}{2}\log(|\bSigma|)-\frac{1}{2}\left[tr\big(\widehat{\Z\Z^{\top}}^{(k)}{\bSigma}^{-1}\big)-
2\widehat{\Z^{\top}}^{(k)}{\bSigma}^{-1}\X\widehat{\bbeta}^{(k+1)}
\right.\right.\nonumber\\
&+& \left.\left.\widehat{\bbeta}^{\top(k+1)}\X^{\top}\bSigma^{-1}\X\widehat{\bbeta}^{(k+1)}\right]\right),\label{CMrhitau}
\end{eqnarray}

\noindent with $\balpha=(\nu^2,\phi)^{\top}$. Note that $\widehat{\tau}^2$ can be recovered via
$\widehat{\tau}^{2(k+1)}=\widehat{\nu}^{2(k+1)}\widehat{\sigma}^{2(k+1)}$. The CM-step (\ref{CMrhitau}) can be easily accomplished using the \texttt{optim} routine in \texttt{R}. This process is iterated until some absolute distance between two successive evaluations of the actual log-likelihood
$\ell(\btheta)$, such as $|\ell(\widehat{\btheta}^{(k+1)})-\ell(\widehat{\btheta}^{(k)})|$ or
$|\ell(\widehat{\btheta}^{(k+1)})/\ell(\widehat{\btheta}^{(k)})-1|$, becomes small enough.

\subsection{Prediction}

Following \cite{diggle2007springer}, let $\Z_{obs}$ denote a vector of random variables with observed realized values, and let $\Z_{pred}$ denote another random variable whose realized values we would like to predict from the observed values of $\Z_{obs}$. A prediction for $\Z_{pred}$ could be any function of $\Z_{obs}$, which we denote by $\hat{\Z}_{pred} = t(\Z_{obs})$. The mean square error (MSE) of $\hat{\Z}_{pred}$ is given by $MSE(\hat{\Z}_{pred})=E((\hat{\Z}_{pred}-\Z_{pred})^2)$; so we can use $\hat{\Z}_{pred}= E(\Z_{pred}|\Z_{obs})$ to find the predictor that minimizes $MSE(\Z_{pred})$.

For a Gaussian process, we write $\Z_{pred}=(Z_{pred}(\mathbf{s}_1),\ldots,Z_{pred}(\mathbf{s}_n))$ for
the unobserved values of the signal at the sampling locations $\mathbf{s}_1,\ldots,\mathbf{s}_n$. We want to predict the signal value at an
arbitrary location; thus our target for prediction is $Z_{pred}(\mathbf{s})$. Given that $(\Z_{obs},\Z_{pred})$ is also a multivariate Gaussian process, we can use the results in Subsection \ref{sec:loglik} to minimize the $MSE(\hat{\Z}_{pred})$, i.e., if $\X^*=(\X_{obs},\X_{pred})$ is the  $(n_{obs}+n_{pred})\times p$ design matrix corresponding to $\Z^*=(\Z_{obs}^\top,\Z_{pred}^\top)$, then $\Z^*\sim N_{n_{obs}+n_{pred}}\left(\X^*\bbeta,\bSigma\right)$, where
$$\bSigma=\begin{bmatrix}
\bSigma_{obs,obs} & \bSigma_{obs,pred}\\
\bSigma_{pred,obs} & \bSigma_{pred,pred}
\end{bmatrix}$$
and $\Z_{pred}|\Z_{obs} \sim N_{n_{pred}}(\bmu_{p},\bSigma_{p})$, where
\begin{equation}\label{mudep}
  \bmu_{p}=\X_{pred}\bbeta + \bSigma_{obs,pred}(\bSigma_{obs,obs})^{-1}(\Z_{obs}-\X_{obs}\bbeta)
\end{equation}
\indent and
$$\bSigma_{p}=\bSigma_{pred,pred}-\bSigma_{pred,obs}(\bSigma_{obs,obs})^{-1}\bSigma_{obs,pred}.$$

\noindent Then, the predictor that minimizes the $MSE$ of prediction will be the conditional expectation in (\ref{mudep}).

\subsubsection{Prediction methods}

Using the previous results, we will now describe the algorithm for each of the three prediction methods that are implemented in this paper.

\begin{enumerate}[label=(\alph*)]

\item \textit{Naive 1 and Naive 2 algorithms}
\vspace{0.2 cm}

We proceed as follows:

\begin{enumerate}[label={\arabic*.}]

\item Impute the censored observations using $LOD$ (called Naive 1 algorithm), or $LOD/2$ (called Naive 2 algorithm), depending on the type of censoring (left or right).

\item Compute the least squares or likelihood estimates for the mean (linear trend in our case) and covariance structure using the imputed data.

\item Evaluate the estimate $\hat{\btheta}$ from step 2 in expression $(\ref{mudep})$ to obtain the predicted values.\\
\end{enumerate}

\item \textit{Seminaive algorithm}
\vspace{0.2 cm}

Here, we follow the method in \cite{Schelin}. Let $C_i$ denote the indicator variable, indicating the presence ($C_i=1$) or absence ($C_i=0$) of a censored observation. Then we proceed as follows:

\begin{enumerate}[label={\arabic*.}]

\item Set $\hat{\Z}_0=(\Z_{obs}^T,\hat{\mathbf{v}}_{0}^T)$ where $\hat{\mathbf{v}}_{0i}= 0$ if $C_i=1$.

\item Obtain $\hat{\btheta}_0=(\hat{\bbeta}, \hat{\sigma^2}, \hat{\phi})$ from $\hat{\Z}_0$ by least squares.

\item Set $k = 0$. Let $\hat{\Z}^{-i}_{k}$ denote the data vector $\hat{\Z}_{k}$  where unit $i$ is removed.

\item Find the predictor $\hat{\Z}(h,\hat{\btheta_k})$ for all $i$ such that $C_i = 1$ using expression (\ref{mudep}) and $\hat{\Z}^{-i}_{k}$, where $h$ denotes the distance between coordinates (isotropy).

\item Set $\hat{\Z}_{k+1}=(\Z_{obs}^T,\hat{\mathbf{v}}_{k+1}^T)$ where $\hat{\mathbf{v}}_{k+1}= (\max(0, \min(\hat{\Z}(h,\hat{\btheta_k}), LOD )) : C_i = 1)$.

\item Update $\btheta_k$ from the new data $\hat{\Z}_{k+1}$.

\item For iteration $k+1$, repeat steps 4-6 until the convergence criterion is satisfied for some constants $c_1$, $c_2$, and $c_3$ chosen by the user.

\item Once convergence is attained, evaluate the estimate $\hat{\btheta}$ in expression $(\ref{mudep})$ to obtain the predicted values.
\end{enumerate}
Note that this algorithm only works for the left-censored case. The constants $c_1$, $c_2$, and $c_3$ are chosen such that the following three conditions are satisfied:
\begin{eqnarray}
\left|\frac{\hat{\sigma^2}(\hat{\Z}_{k+1})-\hat{\sigma^2}(\hat{\Z}_{k})}{\hat{\sigma^2}(\hat{\Z}_{k})}\right|\leq c_1 ,& \hat{\sigma^2}(\hat{\Z}_{k+1}) \leq c_2\hat{\sigma^2}(\Z_{obs}) \mbox{\quad and} & \hat{\xi}{(\hat{\Z}_{k+1}}) >c_3\hat{\xi}{(\Z_{obs})}
\end{eqnarray}

where $\xi(X)$ represents the skewness of $X$.\\

\item \textit{SAEM algorithm}

\vspace{0.2 cm}
Here, we proceed as follows:

\begin{enumerate} [label={\arabic*.}]
\item Obtain the estimates of the mean and covariance structure by SAEM procedure.  For further details, see \cite{ordonez2018geostatistical}.

\item Impute the censored observations by the approximate first moments $\hat{\Z}^{(k)}$, obtained from the SAEM procedure.

\item Evaluate the above estimates in expression (\ref{mudep}) to obtain predicted values for the unobserved locations. In this case, we use $\Z_{obs^{*}}$ to denote the observed data and estimate the censored observations via SAEM. This way, we can easily differentiate it from the fully observed response $\Z_{obs}$.
\end{enumerate}

\end{enumerate}

\subsection{Application: Estimation and prediction tools}

Now, we exemplify the estimation and prediction methods described previously and  implemented in the \texttt{CensSpatial} package. We use the rainfall dataset from Parana state (Brazil) analyzed in \cite{ribeiro}. We carried out some performance comparison for the four prediction methods by using a cross-validation study with the first $100$ observations for estimation and the remaining $43$ observations for comparison of the predictive power of the estimated methods.

The rainfall dataset does not contain censored data, so left censoring was generated artificially for the first $100$ observations, following \cite{Schelin}, i.e., we fixed the censoring level at $\alpha$, sorting the data (in ascending order) and setting the LOD to the 100$\alpha\%$ percentile of the generated sample, i.e., the smallest $\alpha n$ value of $Z(\mathbf{s}_1),\ldots, Z(\mathbf{s}_n)$. This dataset, called \texttt{paranacens25}, is available in the package \texttt{CensSpatial} with 25$\%$ censoring level. The following instructions call the data, then observations that will be used for estimation are stored in the object \texttt{dataest} and those that will be used for prediction are stored in the object \texttt{datapred}.\

\begin{verbatim}
> data(paranacens25)
> dataest=paranacens25[1:100,]
> datapred=paranacens25[101:143,]
\end{verbatim}

\begin{figure}[!h]
\centering
\subfigure[]{\includegraphics[scale=0.5]{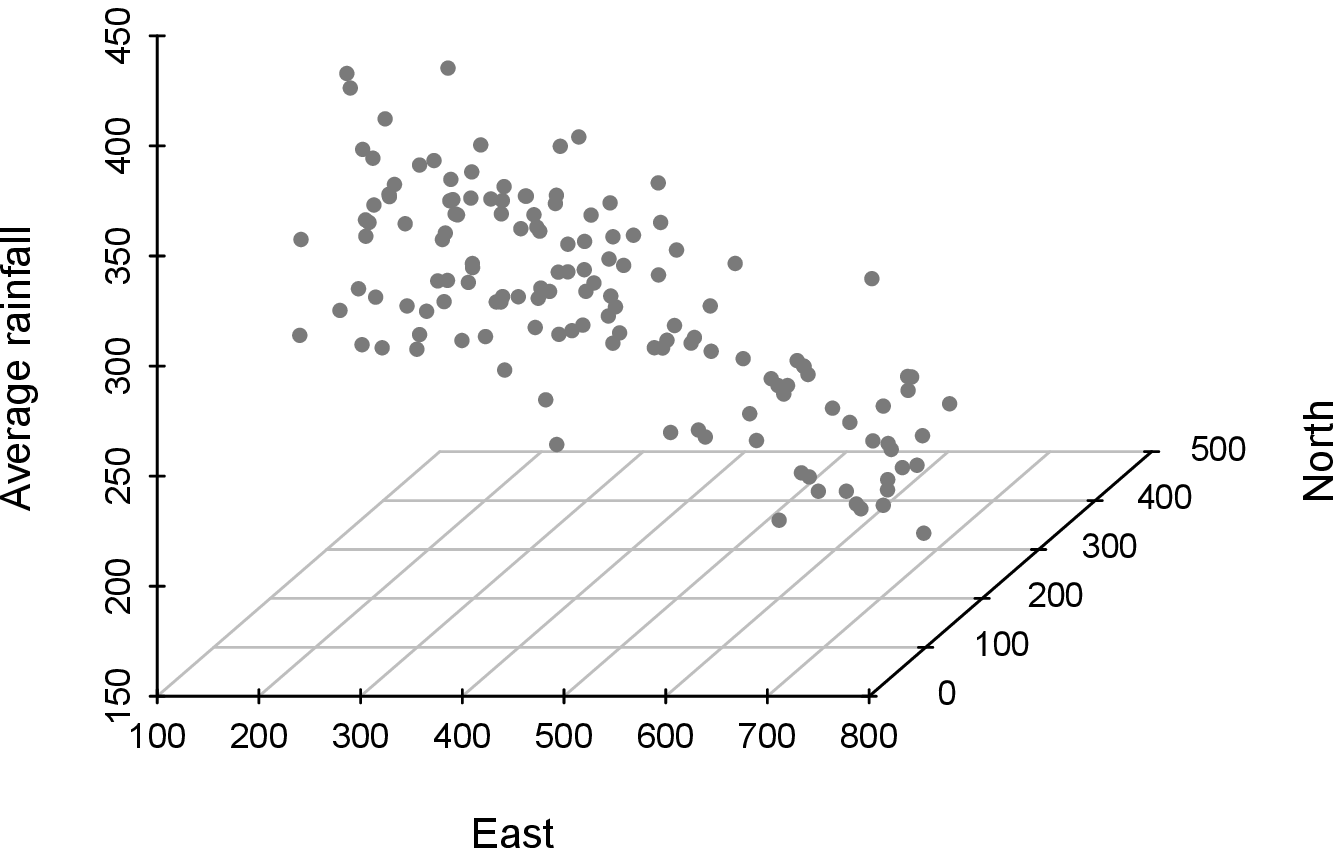}}\hspace{0.8cm}
\subfigure[]{\includegraphics[scale=0.35]{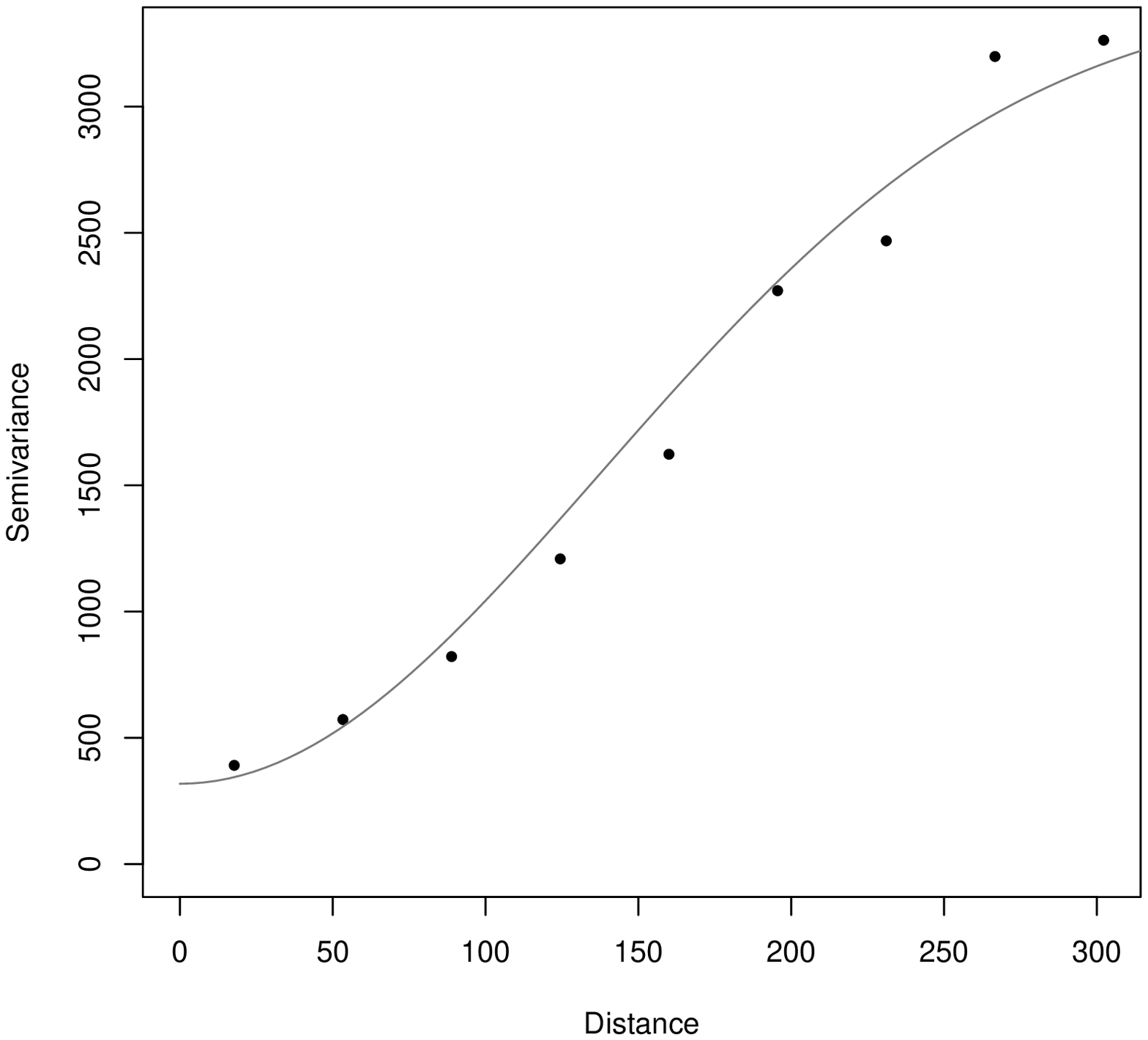}}
\caption{Rainfall dataset: (a) Average rainfall over different locations (b) Sample variogram and Gaussian correlation structure (solid line).} \label{figu1}
\end{figure}

In order to specify the mean and covariance components for the SCL model, we use some descriptive graphs. Figure \ref{figu1} depicts the rainfall dataset, where the left panel shows the average rainfall in each coordinate and the right panel shows the sample variogram for the observed data considering a Gaussian covariance structure. From this figure, we can conclude that it is plausible to assume a linear trend of the rainfall with respect to its coordinates. The behavior showed in the sample variogram seems to indicate that it is also plausible to consider a Gaussian covariance structure for the stochastic component. As an additional exercise, we compare Mat\'{e}rn covariance structures with different $\kappa$ and the Gaussian structure. All these fits present similar results, but we choose the second one because it is simpler and has an easier interpretation.

We fitted the left SCL model defined in (\ref{linearSPM})-(\ref{linearSPM2}) with the specifications previously discussed in the descriptive analysis. To carry out the fit, we used the object \texttt{dataest} defined previously and the four estimation methods available in the \texttt{CensSpatial} package: Seminaive, Naive 1 Naive 2 and SAEM  algorithms. For the Seminaive algorithm, we chose the tuning parameter as $a_1 =0.1$, $a_2 = 2$ and $a_3 = 5$ for the convergence criterion as suggested by \cite{Schelin}. For the SAEM procedure, we performed a Monte Carlo simulation with a sample of size $M = 15$, a maximum number of iterations $W = 200$ and a cutoff point $c = 0.2$. We used the functions \texttt{variofit} and \texttt{likfit} from the \texttt{geoR} package to define the initial values for the SAEM algorithm. These values were used for the remaining algorithms in order to maintain the same conditions in the comparison of the four methods.\\

The estimation process via SAEM can be performed using the \texttt{SAEMSCL} function. Now, we detail how to call the routine and show a print summary of the results obtained through the \texttt{est} object. Arguments of the function are organized in such a way that the first line contains the parameters related to the censored observations, the second one the covariance structure, the third and fourth ones the model specification and finally the last one refers to the SAEM settings. The code below also provides a print summary with the parameter estimates, the linear trend, the covariance structure considered, some information criteria and specific details about the SAEM procedure. Finally, the output for this function is a list with all the information contained in the print summary and specific information of the estimation procedure as the estimates obtained in each iteration. The SAEM parameters for this function must not be changed unless the user knows how they work. We refer to \cite{thais} for details.\

\begin{verbatim}
> cov.ini=c(900,68); y=dataest$y1; cc=dataest$cc
> coords=dataest[,1:2]; coordspred=datapred[,1:2]

> est=SAEMSCL(cc,y,cens.type="left",trend="1st",coords=coords,
+ kappa=2, cov.model="gaussian", fix.nugget=F,nugget=170,
+ inits.sigmae=cov.ini[2],inits.phi=cov.ini[1], search=T,
+ lower=c(0.0001,0.0001),upper=c(50000,5000),
+ M=15,perc=0.25,MaxIter=200,pc=0.2)
\end{verbatim}

\begin{verbatim}
-------------------------------------------------------------------------
  Spatial Censored Linear regression with Normal errors (SAEM estimation)
-------------------------------------------------------------------------

*Type of trend: Linear function of its coordinates,
mu = beta0 + beta1*CoordX + beta2*CoordY

*Covariance structure: Gaussian
---------
Estimates
---------
       Estimated
beta 0  367.1364
beta 1   -0.0758
beta 2   -0.3124
sigma2  832.6448
phi     211.2341
tau2    360.1578
------------------------
Model selection criteria
------------------------
        Loglik     AIC     BIC AICcorr
Value -353.601 719.202 734.833 720.105
-------
Details
-------
Type of censoring = left
Convergence reached? = FALSE
Iterations = 200 / 200
MC sample = 15
Cut point = 0.2
\end{verbatim}

For the Seminaive, Naive 1 and Naive 2 procedures, the estimation process can be carried out using the functions \texttt{Seminaive} and \texttt{algnaive12} ( Naive 1 and Naive 2 algorithms) respectively. As for the \texttt{SAEMSCL} routine, the arguments are organized in such a way that parameters \texttt{data}, \texttt{y.col}, \texttt{coords.col}, \texttt{covar}, \texttt{cc} and \texttt{trend} are related to the censored data structure; parameters \texttt{cov.model}, \texttt{fix.nugget}, \texttt{nugget} and \texttt{kappa} refer to the covariance structure and \texttt{thetaini}, \texttt{cons} and \texttt{MaxIter} refer to the algorithm specifications as initial values and stopping criterion.  The print summary and the output for these functions are quite similar to those offered by the \texttt{SAEMSCL} routine. For our example, the output generated by this function was saved in the objects \texttt{r} and \texttt{s}  for Seminaive and Naive 1, Naive 2 algorithms, respectively, as can be seen in the following instructions: \

\begin{verbatim}
> r=Seminaive(data=dataest,y.col=3,coords.col=1:2,covar=F,cc=cc,trend="1st",
+ cov.model="gaussian",fix.nugget=F,nugget=170,kappa=2,
+ thetaini=c(900,68),cons=c(0.1,2,0.5),MaxIter=200,copred=coordspred)

> s=algnaive12(data=dataest,y.col=3,coords.col=1:2,covar=F,cc=cc,trend="1st",
+ cov.model="gaussian",fix.nugget=F,nugget=170,kappa=2,
+ thetaini=c(900,68),copred=coordspred)

\end{verbatim}

Table \ref{tabi1} shows the estimation results obtained through the three methods. The \texttt{SAEMSCL} method (in boldface) produced the best fit in terms of AIC and BIC values. The second best fitted method was Naive 1.\\

% Table generated by Excel2LaTeX from sheet 'Hoja1'

\begin{table}[!h]
  \centering
  \caption{Rainfall dataset. Models fitted for each spatial estimation and prediction method}
  \scalebox{0.9}{
    \begin{tabular}{lccccccccc}
    \hline
    Method & $\widehat{\beta}_0$ & $\widehat{\beta}_1$ & $\widehat{\beta}_2$ & $\widehat{\sigma}^2$ & $\widehat{\phi}$ & $\widehat{\tau}^2$ & $Loglik$ & $AIC$ & $BIC$ \\
    \hline
    Seminaive &  354.8862 & -0.0710 & -0.2261 & 771.4669 & 198.1979 &  277.8797 &  -439.38 & 890.76 & 906.391\\
    Naive1 & 354.2134 & -0.0671 & -0.2383 & 878.7536 & 119.5233 & 232.2021 & -438.905 & 889.811 & 905.442 \\
    Naive2 & 452.3357 &  -0.2091 & -0.5186 & 1969.1575 & 49.1863 & 579.2321& -501.812 & 1015.625 & 1031.256 \\
   \textbf{SAEM} & \textbf{367.1364} & \textbf{-0.0758} & \textbf{-0.3124} & \textbf{832.6448} & \textbf{211.2341} & \textbf{360.1578} & \textbf{-353.601} &  \textbf{719.202} & \textbf{734.833} \\
    \hline
    \end{tabular}%
    }
  \label{tabi1}%
\end{table}%

Now we proceed to make predictions under each algorithm to compare their performance. To obtain predictions via SAEM, we call the \texttt{predSCL} function, which has three arguments: \texttt{xpred}, which indicates the covariate values at prediction locations; the prediction locations \texttt{coordspred}; and \texttt{est}, which is an object of the class \texttt{"SAEMSpatialCens"} and it is returned by the \texttt{SAEMSCL} function. This routine provides a list with the predicted values along with their standard deviations. The following instructions save the output of this function in the object \texttt{h} for the rainfall averages case. The prediction coordinates were specified using the object \texttt{coordspred} defined previously.

\begin{verbatim}
> xpred=cbind(1,coordspred)
> h=predSCL(xpred=xpred,coordspred = coordspred,est = est)
\end{verbatim}

The functions \texttt{Seminaive} and \texttt{algnaive12} already provide the predicted values for a set of locations \texttt{copred} of interest. For the rainfall average data, if \texttt{r} and \texttt{s} are objects of type \texttt{Seminaive} and \texttt{algnaive12}, the predicted values are provided as \texttt{r\$predictions}, \texttt{s\$predictions1} and \texttt{s\$predictions2}.

\begin{figure}[!h]
\centering
\subfigure[]{\includegraphics[scale=0.48]{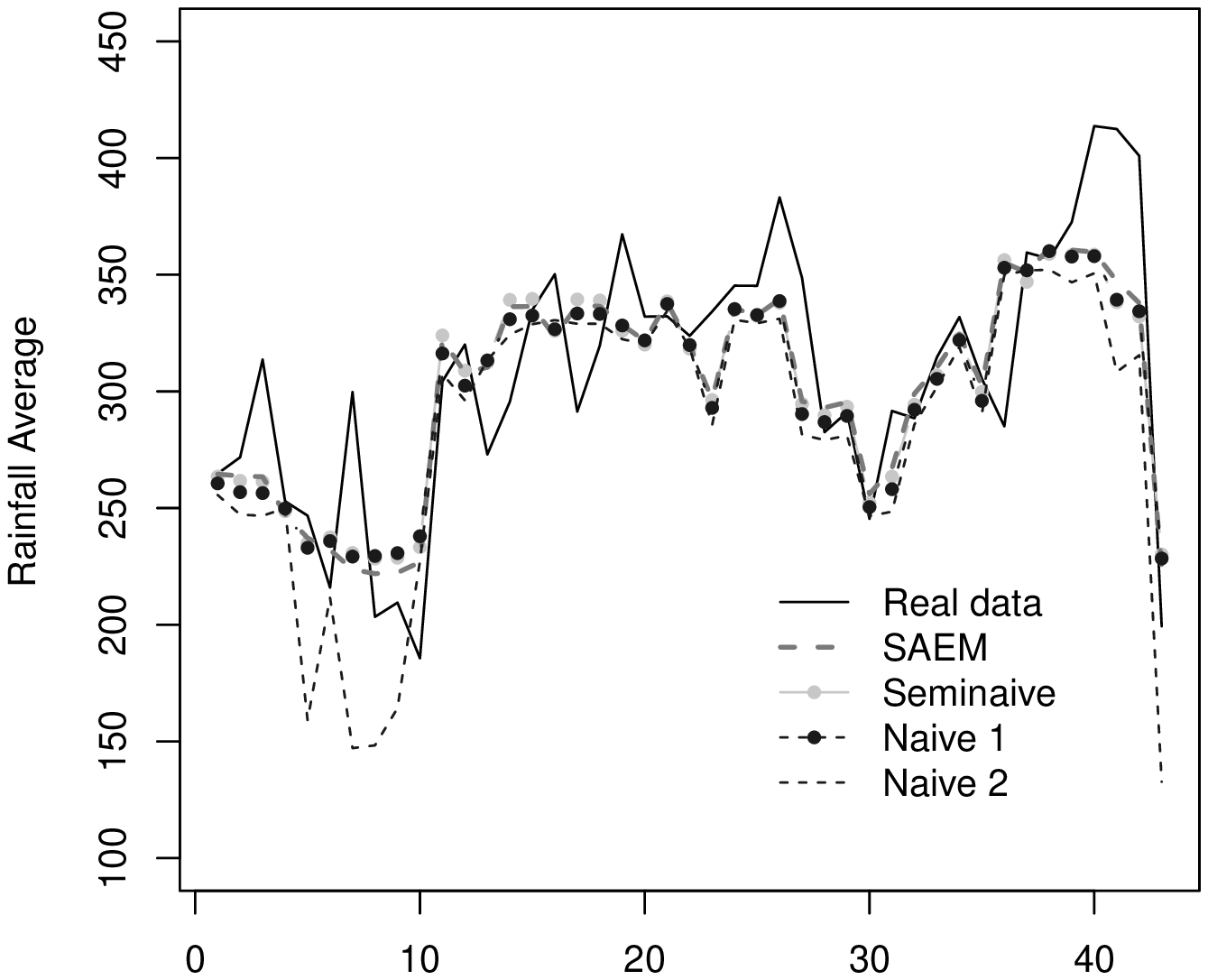}}\hspace{0.5cm}
\subfigure[]{\includegraphics[scale=0.48]{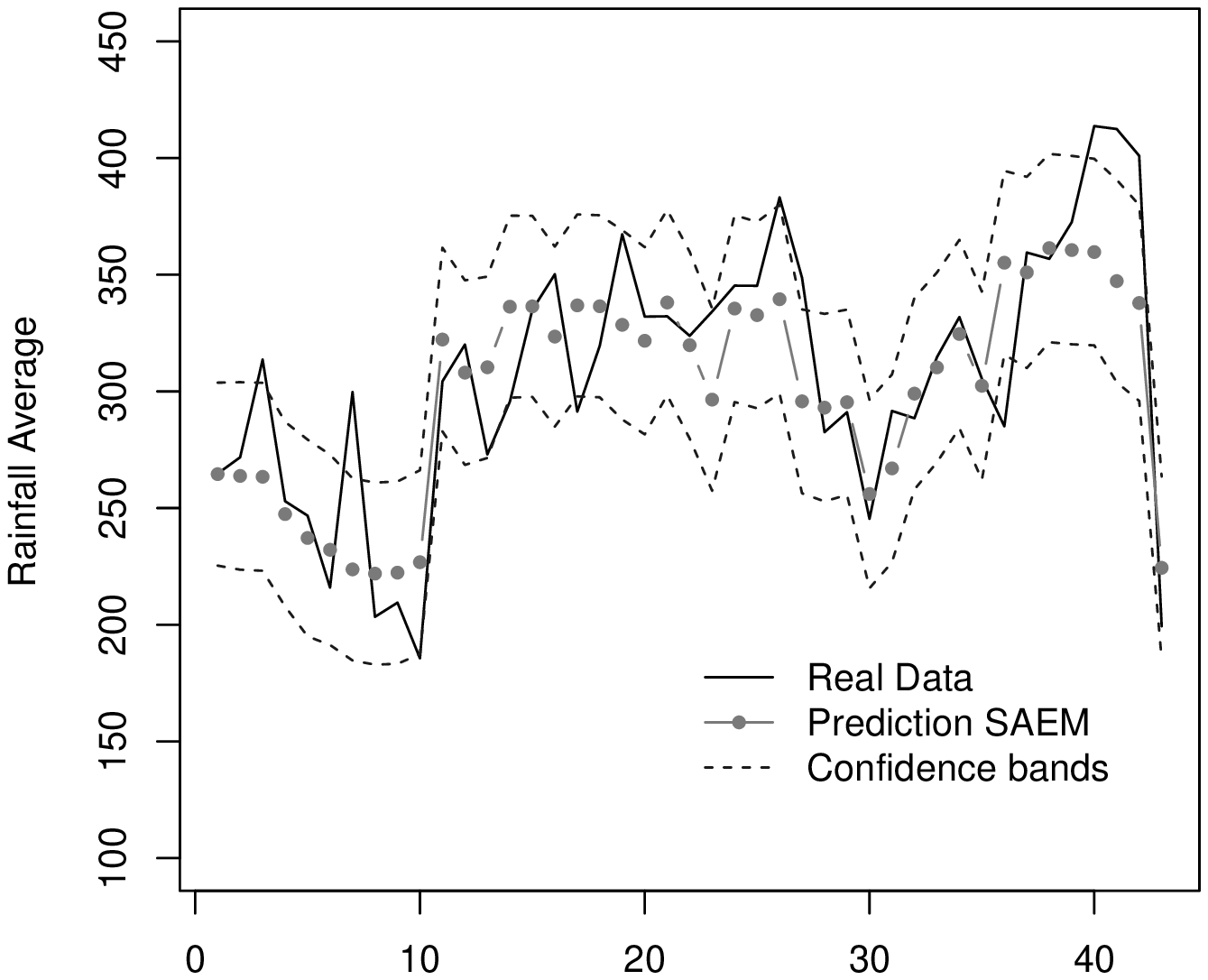}}
\caption{Rainfall dataset: (a) Spatial predictions at the observed locations (b) Real data and a 95\% confidence interval for the prediction via SAEM method.} \label{figu2}
\end{figure}

We can use the results given by these functions to construct graphs that facilitate the prediction interpretation, e.g., panel (b) of Figure \ref{figu2} was plotted using the output of the \texttt{predSCL} function following the next code lines:

\begin{verbatim}
> reval=datapred$y1; predsaem= h$prediction; sdsaem=h$sdpred
> LI=predsaem - (1.96*sdsaem); LS=predsaem + (1.96*sdsaem)

> plot(reval,type="l",ylim=c(100,450),ylab="Rainfall Average",xlab="")
> lines(predsaem,col="gray48",type="both",pch=20)
> lines(LI,col="gray10",lty=2)
> lines(LS,col="gray10",lty=2)
> legend(locator(1),c("Real Data", "Prediction SAEM", "Confidence bands"),
+ col=c(1,"gray48","gray10"),lty=c(1,1,2),pch=c(NA,20,NA),bty="n")
\end{verbatim}

Figure \ref{figu2} shows the predicted values at the observed locations for the four methods. In the left panel (a) we can compare the methods with the true observed values in solid lines. All methods seem to follow the data properly. The right panel (b) shows the real data and a 95$\%$ confidence interval for the prediction obtained via SAEM.

Predictions for the SAEM algorithm may also be assessed graphically using the \texttt{predgraphics} function. This function has \texttt{grid1} as argument, which consists of a grid of coordinates to construct the plot for prediction, \texttt{xpred}, which are the covariate values defined for this grid, \texttt{est} is an object of the class \texttt{"SAEMSpatialCens"} and  \texttt{sdgraph} is a logical argument which is \texttt{TRUE} by default, providing the standard deviation of the predictors. The following instructions illustrate how to use this routine (\texttt{predgraphics}) for the rainfall data.

\begin{verbatim}
> coorgra1=seq(min(coords[,1]),max(coords[,1]),length=50)
> coorgra2=seq(min(coords[,2]),max(coords[,2]),length=50)

> grid1=expand.grid(x=coorgra1,y=coorgra2)
> xpred=cbind(1,grid1)

> graf=predgraphics(xpred=xpred,est=est,grid1=grid1,points=T,sdgraph=T,
+ colors = gray.colors(300),obspoints = sample(1:sum(cc==0),50),main2="")

\end{verbatim}

Figure 3 is generated by the \texttt{predgraphics} function, i.e., two intensity plots, one for the predictors and another for the standard deviation. It also has options for including observed points for comparison purposes.

%
%\begin{verbatim}
%predgraphics(xpred=xpred,est=est,grid1=grid1,points=T,sdgraph=T,
%,obspoints = sample(1:sum(cc==0),50),main2="",main1="")
%\end{verbatim}

\begin{figure}[!h]
\centering
\subfigure[]{\includegraphics[scale=0.4]{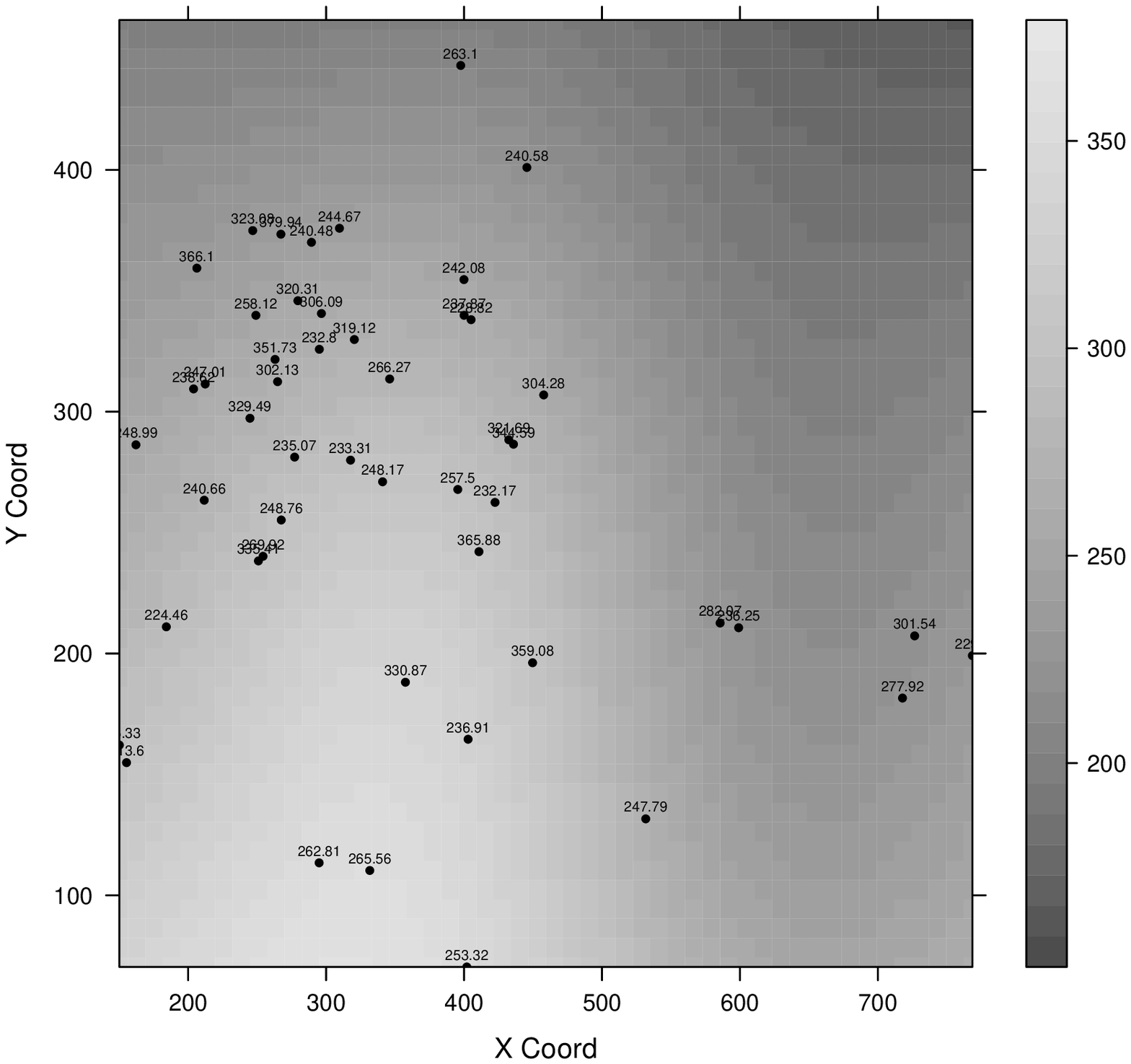}}\hspace{0.5cm}
\subfigure[]{\includegraphics[scale=0.4]{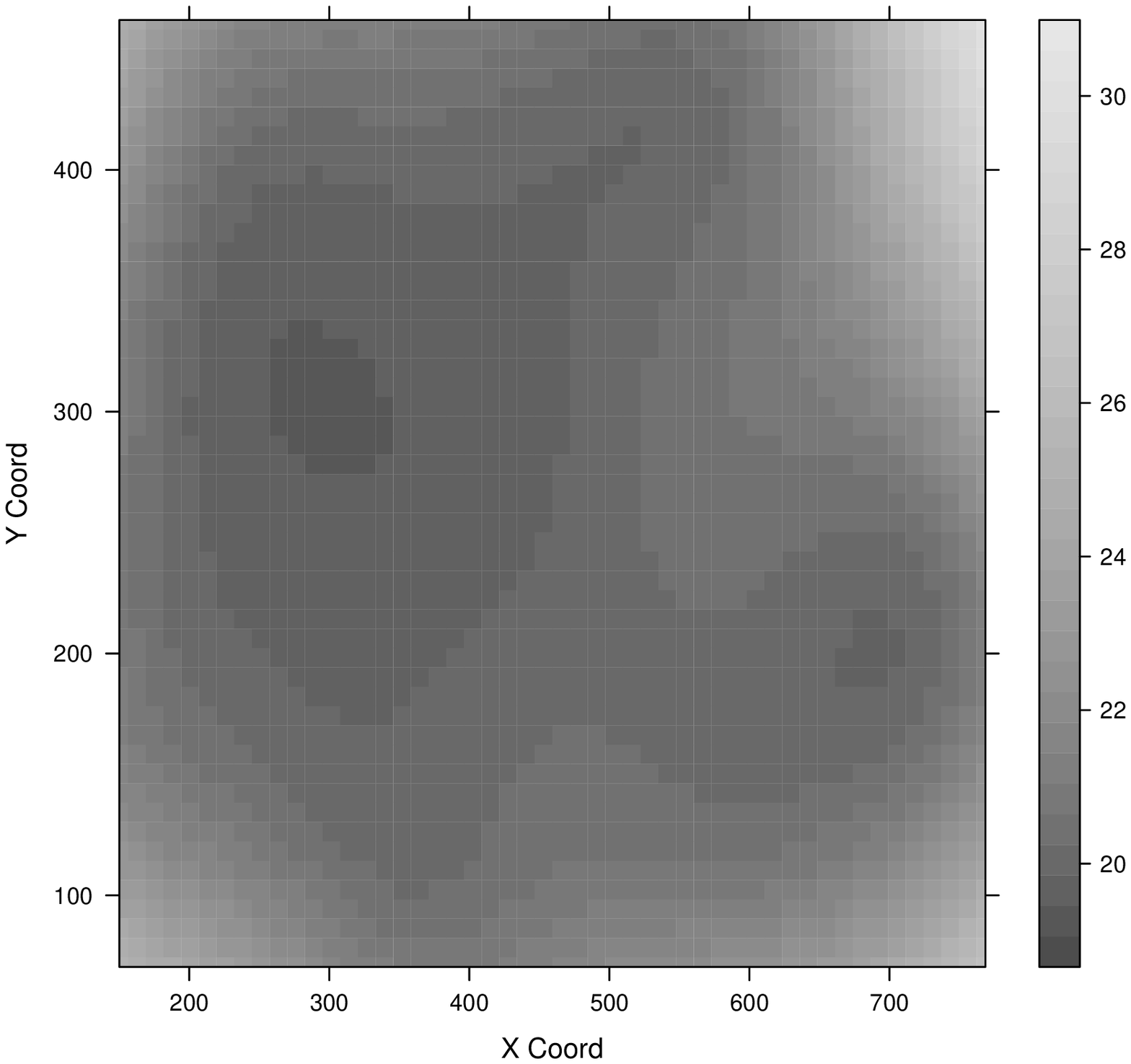}}
\caption{Rainfall dataset: (a) Observed values and intensity plot for the predicted values (b) Intensity plot for the standard deviation of prediction via SAEM method.} \label{figu3}
\end{figure}

We used the mean square prediction error (MSPE) as a measure of prediction power. This is defined as \citep[see,][]{fridley2006data}:
\begin{eqnarray}\label{MSE}
  MSPE &=& \frac{1}{n}\sum_{i=1}^n (Z_i-\hat{Z}_i)^2,
\end{eqnarray}

where $Z_i$ is the observed value, $\hat{Z}_i$ is the predicted value and $n$ is the number of samples predicted. Since the MSPE is an error measure, lower values lead to better models. Thus, using the ${MSEP}$ we can also conclude that the SAEM method presents the best fit, with a $\sqrt{MSEP}$ equal to $5.649$, compared with its competitors, all with $\sqrt{MSEP}$ greater than $5.9$.

\section{Local influence diagnostic utilities}\label{s2}

In this section, we describe all measures used for local diagnostics. All calculations are based in the Hessian matrix of log-likelihood conditional expectation and the SAEM estimates obtained through the \texttt{SAEMSCL} routine. We used two simulated processes to illustrate how the package works to detect influential observations.

Following \cite{ZhuLee2001}, we compute the Hessian matrix $\tfrac{\partial^2 Q(\btheta, \bomega|\hat{\btheta})}{\partial \btheta \partial \btheta^{\top}}$ in order to obtain local diagnostic measures for a particular perturbation scheme as described in later subsections. Let $\btheta= (\bbeta^{\top},\balpha{^\top})^{\top}$ with $\balpha=(\alpha_1,\alpha_2,\alpha_3)^{\top}$ where $\alpha_1=\sigma^2$, $\alpha_2=\phi$ and $\alpha_3=\tau^2$. Then, the necessary expressions to compute the Hessian matrix $\ddot{Q}(\btheta|\hat{\btheta})$ are given by:
%\begin{align*}
%  \frac{\partial^2 Q(\btheta, \bomega|\hat{\btheta}))}{\partial \bbeta \partial \bbeta^{\top}} &= \X^{\top}\bSigma^{-1}\X \\
%  \frac{\partial^2 Q(\btheta, \bomega|\hat{\btheta}))}{\partial \bbeta \partial \alpha_{k}} &= \X^{\top}\frac{\partial \bSigma^{-1}}{\partial \alpha_k}\widehat{\Z}- \X^{\top}\frac{\partial \bSigma^{-1}}{\partial \alpha_k}\X\bbeta \\
%  \frac{\partial^2 Q(\btheta, \bomega|\hat{\btheta}))}{\partial \alpha_k \partial \alpha_l} &= \mathrm{tr}\left(\frac{\partial \bSigma^{-1}}{\partial \alpha_l}\frac{\partial \bSigma^{-1}}{\partial \alpha_k} + \frac{\partial^2 \bSigma^{-1}}{\partial \alpha_k\alpha_l}\bSigma^{-1} \right)- \frac{1}{2}\left[\mathrm{tr}\left(\widehat{\Z\Z}^{\top}\frac{\partial^2 \bSigma^{-1}}{\partial \alpha_k\alpha_l}\right) -2\bbeta^{\top}\X^{\top}\frac{\partial^2 \bSigma^{-1}}{\partial \alpha_k\alpha_l}\widehat{\Z}\right.\\
%  &+ \left.\bbeta^{\top}\X^{\top}\frac{\partial^2 \bSigma^{-1}}{\partial \alpha_k\alpha_l}\X\bbeta\right]
%\end{align*}
\begin{align}\label{hessian}
  \frac{\partial^2 Q(\btheta, \bomega|\hat{\btheta})}{\partial \bbeta \partial \bbeta^{\top}} &= \X^{\top}\bSigma^{-1}\X, \\
  \frac{\partial^2 Q(\btheta, \bomega|\hat{\btheta})}{\partial \bbeta \partial \alpha_{k}} &= \X^{\top}\frac{\partial \bSigma^{-1}}{\partial \alpha_k}\widehat{\Z}- \X^{\top}\frac{\partial \bSigma^{-1}}{\partial \alpha_k}\X\bbeta, \\
  \frac{\partial^2 Q(\btheta, \bomega|\hat{\btheta})}{\partial \alpha_k \partial \alpha_l} &= \mathrm{tr}\left(\frac{\partial \bSigma^{-1}}{\partial \alpha_l}\frac{\partial \bSigma^{-1}}{\partial \alpha_k} + \frac{\partial^2 \bSigma^{-1}}{\partial \alpha_k\alpha_l}\bSigma^{-1} \right)\\\nonumber
  &- \frac{1}{2}\left[\mathrm{tr}\left(\widehat{\Z\Z}^{\top}\frac{\partial^2 \bSigma^{-1}}{\partial \alpha_k\alpha_l}\right) -2\bbeta^{\top}\X^{\top}\frac{\partial^2 \bSigma^{-1}}{\partial \alpha_k\alpha_l}\widehat{\Z} + \bbeta^{\top}\X^{\top}\frac{\partial^2 \bSigma^{-1}}{\partial \alpha_k\alpha_l}\X\bbeta\right].
\end{align}

\subsection{Local influence}

We derive the normal curvature of the local influence for some common perturbation schemes that could be present either in the model or the data. We will consider perturbations in response, scale matrix and explanatory variables.

Consider a perturbation vector $\bomega$ varying in an open region $\bOmega \subset \mathbb{R}^n$. The conditional log-likelihood expectation of the complete data for the perturbed model will be denoted by $Q(\btheta,\bomega| \hat{\btheta})=E(\ell(\btheta,\bomega|\z_c)|\mathbf{V}, \mathbf{C} ,\hat{\btheta})$ which reaches its maximum at $\hat{\btheta}(\bomega)$. Assume there is an $\bomega_0$ in $\bOmega$ such that $\ell(\btheta,\bomega_0|\z_c)=\ell(\btheta|\z_c)$ for all $\btheta$. We use the $Q$-displacement function defined as:

$$f_Q(\bomega)=2[Q(\btheta|\hat{\btheta})- Q(\btheta|\hat{\btheta}(\bomega))],$$
to obtain the influence graph $\delta(\bomega)=(\bomega^{\top},f_Q(\bomega))^{\top}$. The local behavior of $f_Q(\bomega)$ can be summarized using the curvature $C_{f_{Q,\mathbf{d}}}$ at $\bomega_0$ in the direction of some unit vector $\mathbf{d}$. This curvature is defined by:

$$C_{f_{Q,\mathbf{d}}}=-2\mathbf{d}^{\top}\ddot{Q}_{\bomega_{0}}\mathbf{d}\hspace{0.2cm}\quad \textup{and} \quad\hspace{0.2cm} -\ddot{Q}_{\bomega_0}=\Delta_{\bomega_0}^{\top}-\ddot{Q}(\btheta|\hat{\btheta})^{-1}\Delta_{\bomega_0},$$
where $\Delta_{\bomega}=\frac{\partial^2Q(\btheta,\bomega|\hat{\btheta})}{\partial \btheta \partial \bomega^{\top}}$ and $\ddot{Q}(\btheta|\hat{\btheta})$ is the Hessian matrix with elements as in expressions (\ref{hessian})-(13).

As described in \cite{cook86}, the quantities $C_{f_{Q,\mathbf{d}}}$ and $\ddot{Q}_{\bomega_0}$ are useful for detecting influential observations. Let the spectral (orthonormal) decomposition of $-2\ddot{Q}_{\bomega_0}=\sum_{i=1}^n{\lambda_i\mathbf{a}_i\mathbf{a}_i^{\top}}$ with  $\lambda_1 \geq \lambda_2 \geq \cdots \lambda_r > \lambda_{r+1}=\cdots=\lambda_n=0$ eigenvalues and $\mathbf{a}_i,$ for $i=1,\cdots,n$, eigenvectors of this matrix. \cite{ZhuLee2001} proposed inspecting all eigenvectors corresponding to nonzero eigenvalues to capture the relevant information about influential observations.

Consider the aggregated contribution of each eigenvector with a corresponding nonzero eigenvalue as $\tilde{\lambda_i}=\lambda_i/ (\lambda_1+\cdots + \lambda_r)$. Let $\mathbf{a}_i^2=(a_{11}^2,\cdots,a_{in}^2)^{\top}$ and $M(0)=\sum_{i=1}^{r} \lambda_i\mathbf{a}_i^2$ with $M(0)_l=\sum_{i=1}^{r} \lambda_i\mathbf{a}_{il}^2$ for $l=1,\cdots, n$, the $l$-th component of $M(0)$. The assessment of influential observations can be performed through visual inspection of $M(0)_l$ plotted against the index $l$. The $l$-th case may be considered influential if $M(0)_l$ is larger than a benchmark value. \cite{ZhuLee2001} also considered the conformal normal curvature $B_{f_{Q,\mathbf{d}}}(\btheta)=C_{f_{Q,\mathbf{d}}}(\btheta)/tr[-2\ddot{Q}_{\bomega_0}]$, whose computation is quite simple and has the property that $0 \leq B_{f_{Q,\mathbf{d}}}(\btheta)\leq 1$. They also showed that for a perturbation vector $\mathbf{d}_l$ whose $l$-th entry is 1 and the remaining entries are 0,  $M(0)_l= B_{f_{Q,\mathbf{d}_l}}(\btheta)$, so we can obtain $M(0)_l$ through $B_{f_{Q,\mathbf{d}_l}}(\btheta)$.

In general, there is no rule to judge how influential a specific case in the data is. Some measures have been proposed to determine the benchmark value in the visual inspection of $M(0)$. Let $\bar{M}(0)$ and $SM(0)$ be the mean and standard error of the vector $M(0)$ respectively. \cite{poon1999conformal} proposed using $2\bar{M}(0)$ as a benchmark value while \cite{ZhuLee2001} proposed $\bar{M}(0)+2SM(0)$, for taking into account the standard deviation of $M(0)$. In the \texttt{CensSpatial} package, we use the benchmark proposed by \cite{LeeXu04} as $\bar{M}(0)+c^*SM(0)$, where $c^*$ is a constant, which can vary depending of the specific application.

\subsection{Perturbation schemes}

We consider three types of perturbation schemes: {perturbation in the response variable}, which can indicate influence in their own predicted values, {perturbation in the scale matrix} $\bSigma$, which may reveal the individuals that are influential in the estimation of the scale parameter $\balpha$ and finally {perturbation in explanatory variables}. For all these schemes, we are interested in computing the $\Delta_{\bomega}$ matrix with elements:

$$\Delta_{\bbeta}=\frac{\partial^2Q(\btheta,\bomega|\hat{\btheta})}{\partial \bbeta \partial\bomega^{\top}} \hspace{0.1cm}\quad \textup{and} \quad\hspace{0.1cm} \Delta_{\balpha}= (\Delta_{\alpha_1}^{\top},\Delta_{\alpha_2}^{\top},\Delta_{\alpha_3}^{\top}),$$

where $\Delta_{\alpha_k}=\frac{\partial^2Q(\btheta,\bomega|\hat{\btheta})}{\partial \alpha_{k} \partial \bomega^{\top}}$, for $k=1,2,3$.

\subsubsection{Response perturbation}

A perturbation in response variable $\mathbf{V}$ can be introduced by replacing it with $\mathbf{V}(\bomega)= \mathbf{V}+ \bomega$, recalling the non-perturbed model $V_i=V_{0i}$ for $C_i=0$ and that $V_i$ belongs to $[V_{1i},V_{2i}]$ for $C_i=1$. In this way we can write the perturbed model as:
\vspace{-0.5cm}

\begin{eqnarray}\label{CensL2}
C_i =
\left\{\begin{array}{ccc}
  1 & if & V_{1i}(\omega_i)\leq Z_{i}(\omega_i)\leq V_{2i}(\omega_i)\,, \\
  0 & if & Z_{i}(\omega_i) = V_{0i}(\omega_i)\,,
  \end{array}\right. \nonumber
\end{eqnarray}
where $\mathbf{V}_1(\bomega)=\mathbf{V}_1+ \bomega$, $\mathbf{V}_2(\bomega)=\mathbf{V}_2+\bomega$ and $\Z(\bomega)=\Z-\bomega$. To obtain the perturbed function $Q(\btheta|\hat{\btheta},\bomega)$, we replace $\widehat{\Z}$ and $\widehat{\Z\Z^{\top}}$ with $\widehat{\Z}_{\omega}=\widehat{\Z}- \bomega$ and $\widehat{\Z}_{\omega}\widehat{\Z_{\omega}}^{\top}=\widehat{\Z\Z^{\top}}-\widehat{\Z}\bomega- \bomega\widehat{\Z}^{\top}+\bomega\bomega^{\top}$ in the non-perturbed $Q(\btheta|\hat{\btheta})$ from Subsection 2.3. Let $\bomega_0=\mathbf{0}$ be the vector that represents the case of no perturbation in response. Then, the matrix $\Delta_{\bomega_0}$ has the elements $\Delta_{\bbeta}=-\X^{\top}\bSigma^{-1}$ and $\Delta_{\balpha}=[\Delta_{\alpha_{ki}}],$ where
$$\Delta_{\alpha_{ki}}= \frac{1}{2} \left((\Z_{(i)}+\Z_{(i)}^{\top})\frac{\partial \bSigma^{-1}}{\partial \alpha_k}\right) +
\Z_{(i)}^{\top}\mathbf{P}^{\top}\frac{\partial \bSigma^{-1}}{\partial \alpha_k},$$
with $\mathbf{P}=\X(\X^{\top}\bSigma^{-1}\X)^{-1}\X^{\top}\bSigma^{-1}$ and $\Z_{(i)}$ the matrix with the $i$-th row being $\widehat{\Z}$ and the remaining rows being zero. Expressions for $\tfrac{\partial \bSigma^{-1}}{\partial \alpha_k}$ are defined in the Appendix.

\subsubsection{Scale matrix perturbation}

Now, we consider the scheme perturbation of the form $\bSigma(\bomega)=\mathbf{D}(\bomega)\bSigma$ to study the effects of perturbation on the scale matrix. In this case $\mathbf{D}(\bomega)$ is an $n \times n$ diagonal matrix with elements $\bomega$. The non-perturbed model is obtained when $\omega_i= 1$ for all $i=1,\cdots,n$ and therefore $\Delta_{\bomega_0}$ is a $(p+3)\times n$ matrix with components $\Delta_{\bbeta_i}=\X^{\top}\bSigma^{-1}\mathbf{d}(i)[\mathbf{I_n}-\mathbf{P}]\Z$ and $\Delta_{\balpha}=[\Delta_{\alpha_{ki}}]$, with
\begin{multline*}
\Delta_{\alpha_{ki}}= -\mathrm{tr}\left(\bSigma^{-1}\mathbf{d}(i)\frac{\partial \bSigma}{\partial \alpha_k}\right)\\
-\frac{1}{2}\left[\mathrm{tr}\left(\widehat{\Z\Z^{\top}}\left( \bSigma^{-1}\frac{\partial\bSigma}{\partial \alpha_k}\bSigma^{-1}\right)\mathbf{d}(i)\right)+ \widehat{\Z}^{\top}[\mathbf{P}^{\top}-2\mathbf{I}]\left(\bSigma^{-1}\frac{\partial\bSigma}{\partial \alpha_k}\bSigma^{-1}\right)\mathbf{d}(i)\mathbf{P}\widehat{\Z}\right],
\end{multline*}
for $k=1,2,3$, where $\mathbf{d}(i)$ is an $n \times n$ matrix with the $i$-th diagonal element equal to 1 and the remaining elements equal to zero.

\subsubsection{Explanatory variable perturbation}

We consider for this scheme a perturbation of the form $\X(\bomega)= \mathbf{X}+\mathbf{W}$ and replace the perturbed $Q(\btheta|\hat{\btheta},\bomega)$ function as in the response perturbation, in this case, $\mathbf{W}=\bomega\mathbf{1}^{\top}$. Under the non-perturbed model, $\bomega=\mathbf{0}$ and therefore $\Delta_{\bomega_0}$ has elements $\Delta_{\bbeta_i}= \widehat{\Z}^{\top}[\mathbf{I}-\mathbf{2P^{\top}}]\bSigma^{-1}\mathbf{W_{\textit{\textrm{i}}}^{(1)}}$ and $\Delta_{\balpha}=[\Delta_{\alpha_{ki}}]$, with
$$\Delta_{\alpha_{ki}}= \widehat{\Z}^{\top}[\mathbf{I-P^{\top}}]\frac{\partial \bSigma^{-1}}{\partial \alpha_k}\mathbf{W_{\textit{\textrm{i}}}^{(1)}}(\X^{\top}\bSigma^{-1} \X)^{-1}\X^{\top}\bSigma\widehat{\Z},$$

where $\mathbf{W_{\textit{\textrm{i}}}^{(1)}}$ is an $n \times p$ matrix with the $i$-th row equal to 1 and the remaining rows equal to zero.

%%%%%%%%%%%%%%%%%%%%%%%%%%%%%%%%%%%%%%%%%%%%%%%%%%%%%%%%%%%%%%%%%%%%%%%%%%%%%%%
%%%%%%%%%%%%%%%%%%%%%%%%%%%%%%%%%%%%%%%%%%%%%%%%%%%%%%%%%%%%%%%%%%%%%%%%%%%%%%%

\subsection{Application: Diagnostic tools}

In this subsection, we use simulated data to illustrate the local diagnostic utilities that the  \texttt{CensSpatial} package offers. The dataset corresponds to two linear left censored processes with the same mean structure but with different covariance structures for errors, i.e., we set a process as in equations (\ref{linearSPM})-(\ref{linearSPM2}) with spherical and Mat\'{e}rn ($\kappa=0.3$) covariance structures and two covariates $\X_1 \sim U(0,1)$ and $\X_2 \sim U(0,3)$ for the mean structure. We set a censoring  level of $15\%$ and the observations $\#91$,$\#126$ and $\#162$ as atypical points for the two processes.

To generate the spatial random samples used in the simulations, we call the function \texttt{rspacens}, its principal arguments are \texttt{beta} and \texttt{cov.ini} to fix the parameters of the mean and covariance structure respectively, \texttt{coords} to define the coordinates used for the sample, \texttt{cens} for the censoring level, \texttt{cens.type} to define the censoring type (left or right) and \texttt{cov.model} to define the spatial correlation function to use. The code below illustrates how we generate the samples and the atypical points used in this paper.
\\

\begin{verbatim}
> nobs=200; npred=100
> r1=sample(seq(1,30,length=400),n+n1); r2=sample(seq(1,30,length=400),n+n1)
> coords=cbind(r1,r2); cov.ini=c(2,0.1); beta=c(5,3,1)
> xtot<-cbind(1,runif((n+n1)),runif((n+n1),2,3)); xobs=xtot[1:n,]

> obj=rspacens(cov.pars=c(3,.3,0),beta=beta,x=xtot,coords=coords,cens=0.15,
+ n=(nobs+ npred),n1=npred,cov.model="matern" ,cens.type="left",kappa=0.3)

> y=obj$datare[,3]

> y[91]=y[91]+ 5*sd(y); y[126]=y[126]+ 5*sd(y); y[162]=y[162]+ 5*sd(y)
\end{verbatim}

To assess the influence of outliers in the estimation process, we use the \texttt{localinfmeas} function. This uses an object of class \texttt{"SAEMSpatialCens"}, i.e., an object as a result of the SAEM estimation procedure to compute the value of the vector $M(0)$ for each perturbation scheme and detect the influential points using the \cite{LeeXu04} criteria. The benchmark used for the routine is $c^{*}=3$ by default. On the other hand, the function \texttt{atypical} filters the results to get only the atypical points detected by the \texttt{localinfmeas} routine. The following instructions generate the output of the \texttt{localinfmeas} and \texttt{atypical} functions, considering a linear process with Mat\'{e}rn covariance structure and three perturbation schemes discussed previously.

\begin{verbatim}
> sest=SAEMSCL(cc,y,cens.type="left",trend="other",x=xobs,coords=coords,
+ M=15,perc=0.25,MaxIter=5,pc=0.2,cov.model=type,kappa=0.3,fix.nugget=T,
+ nugget=0,inits.sigmae=cov.ini[1], inits.phi=cov.ini[2],search=T,
+ lower=0.00001,upper=50)

> w = localinfmeas(sest, fix.nugget = TRUE, c = 3)
> atypical(w)$RP

             obs         m0
32  atypical obs 0.04478987
91  atypical obs 0.09086417
126 atypical obs 0.10935255
162 atypical obs 0.09978107
\end{verbatim}

In these lines, \texttt{w} is a \texttt{localinfmeas} object. Then \texttt{atypical(w)} will be a list with three elements: \texttt{EP}, \texttt{SP} and \texttt{RP}, containing the atypical detected observations for each perturbation scheme. By default, \texttt{localinfmeas} also generates a diagnostic plot with for the three perturbation schemes.

\begin{figure}[!h]
\centering
\subfigure[]{\includegraphics[width = 0.325\textwidth]{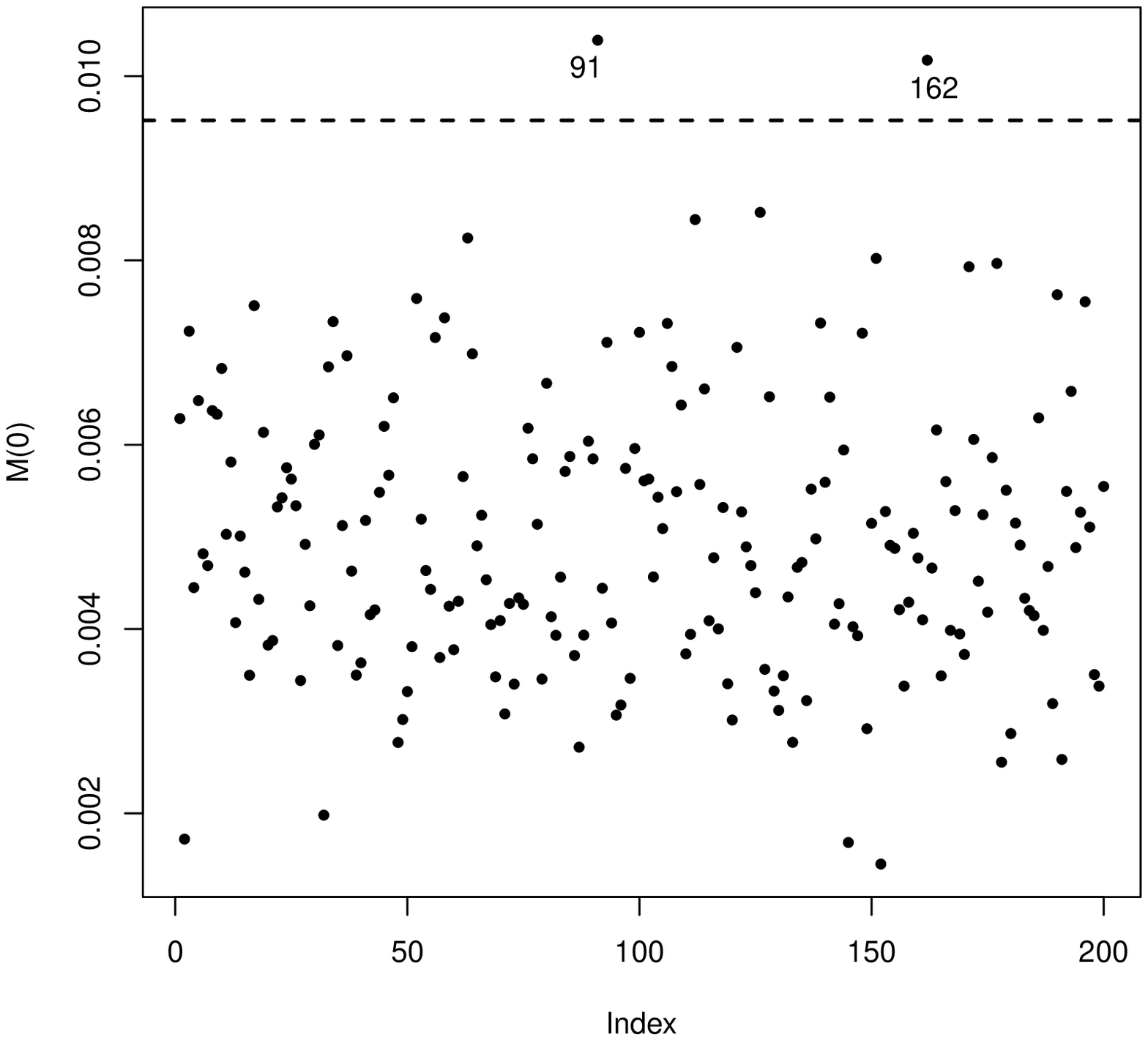}}
\subfigure[]{\includegraphics[width = 0.325\textwidth]{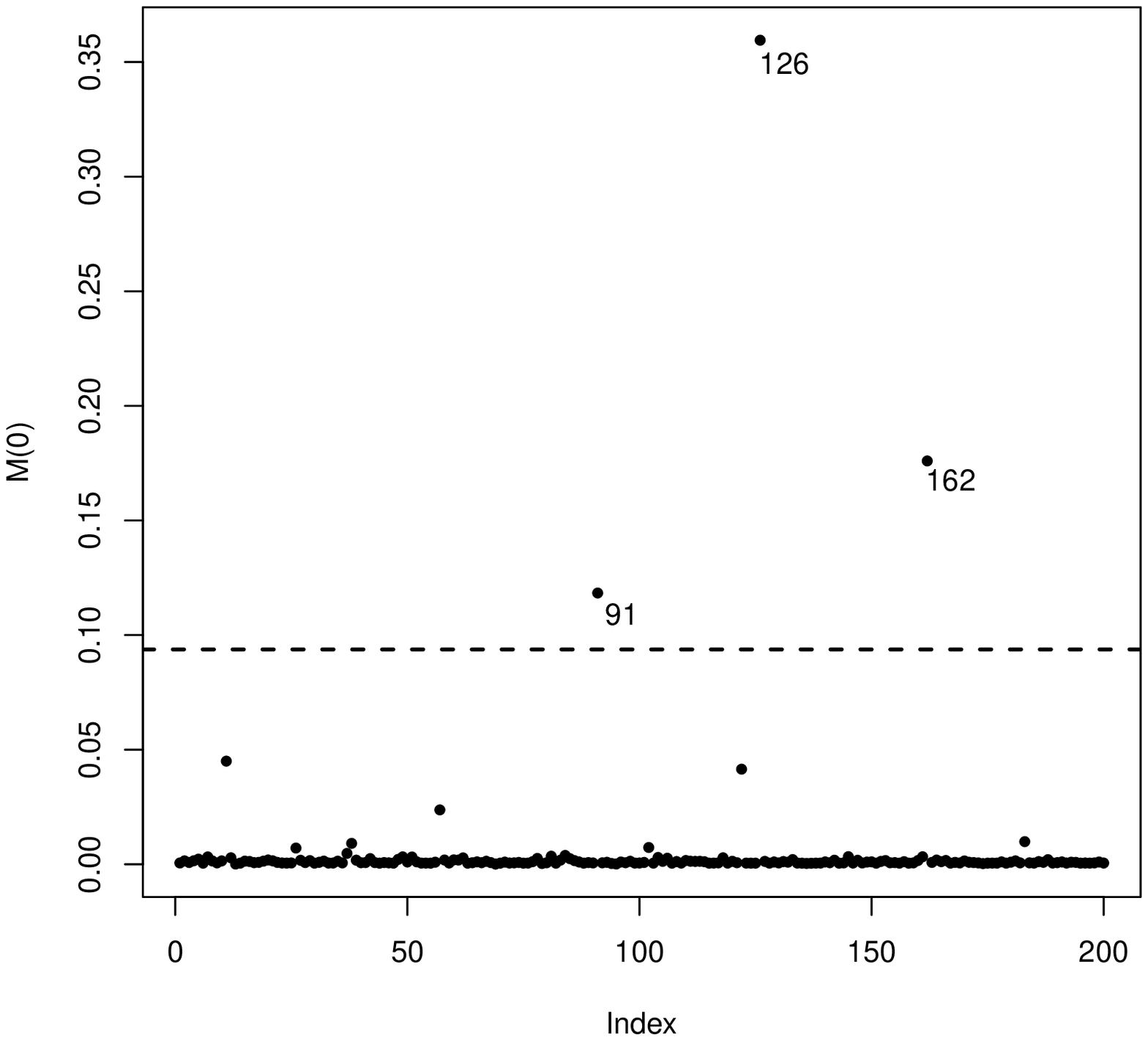}}
\subfigure[]{\includegraphics[width = 0.325\textwidth]{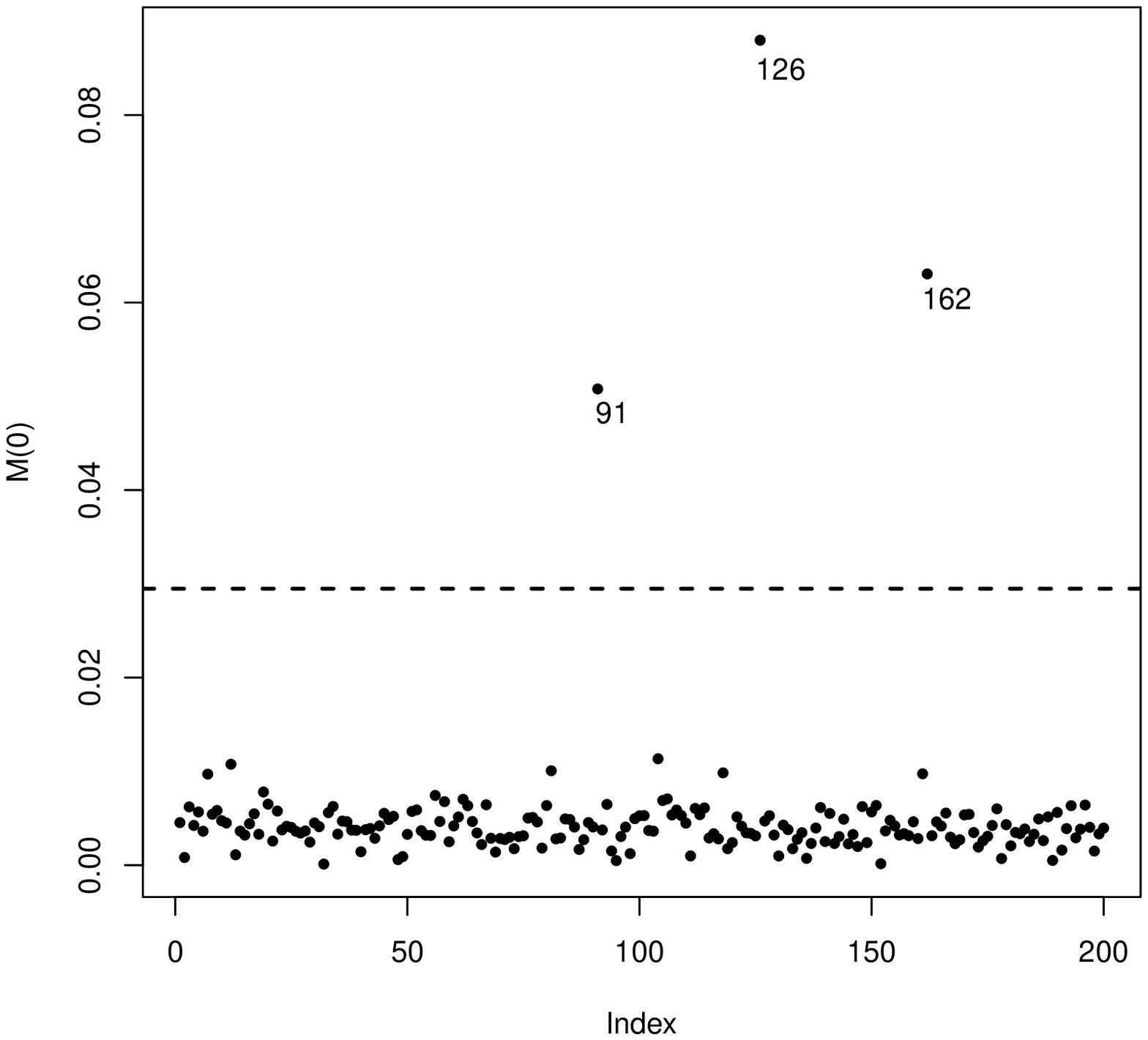}}\\
\subfigure[]{\includegraphics[width = 0.325\textwidth]{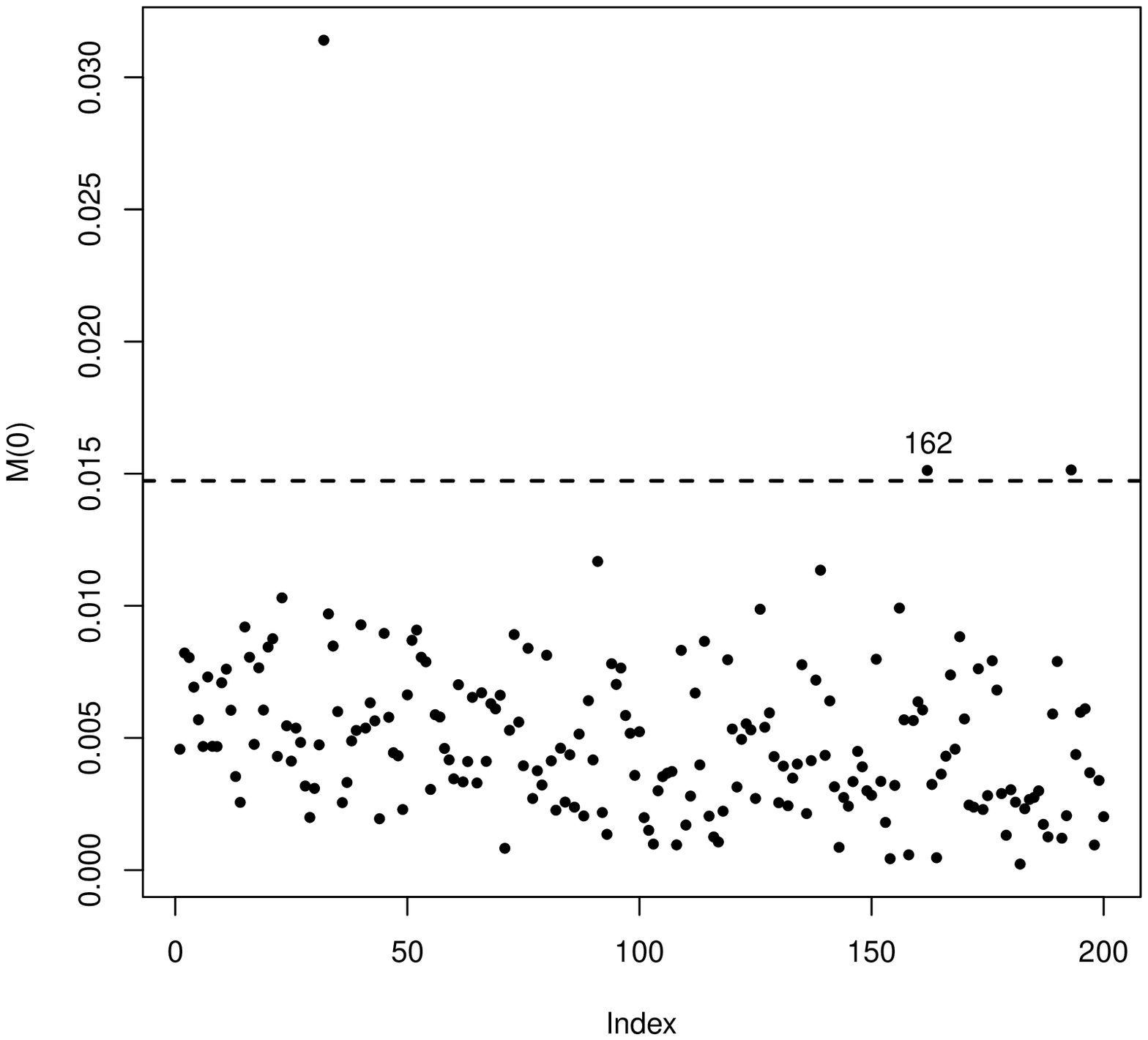}}
\subfigure[]{\includegraphics[width = 0.325\textwidth]{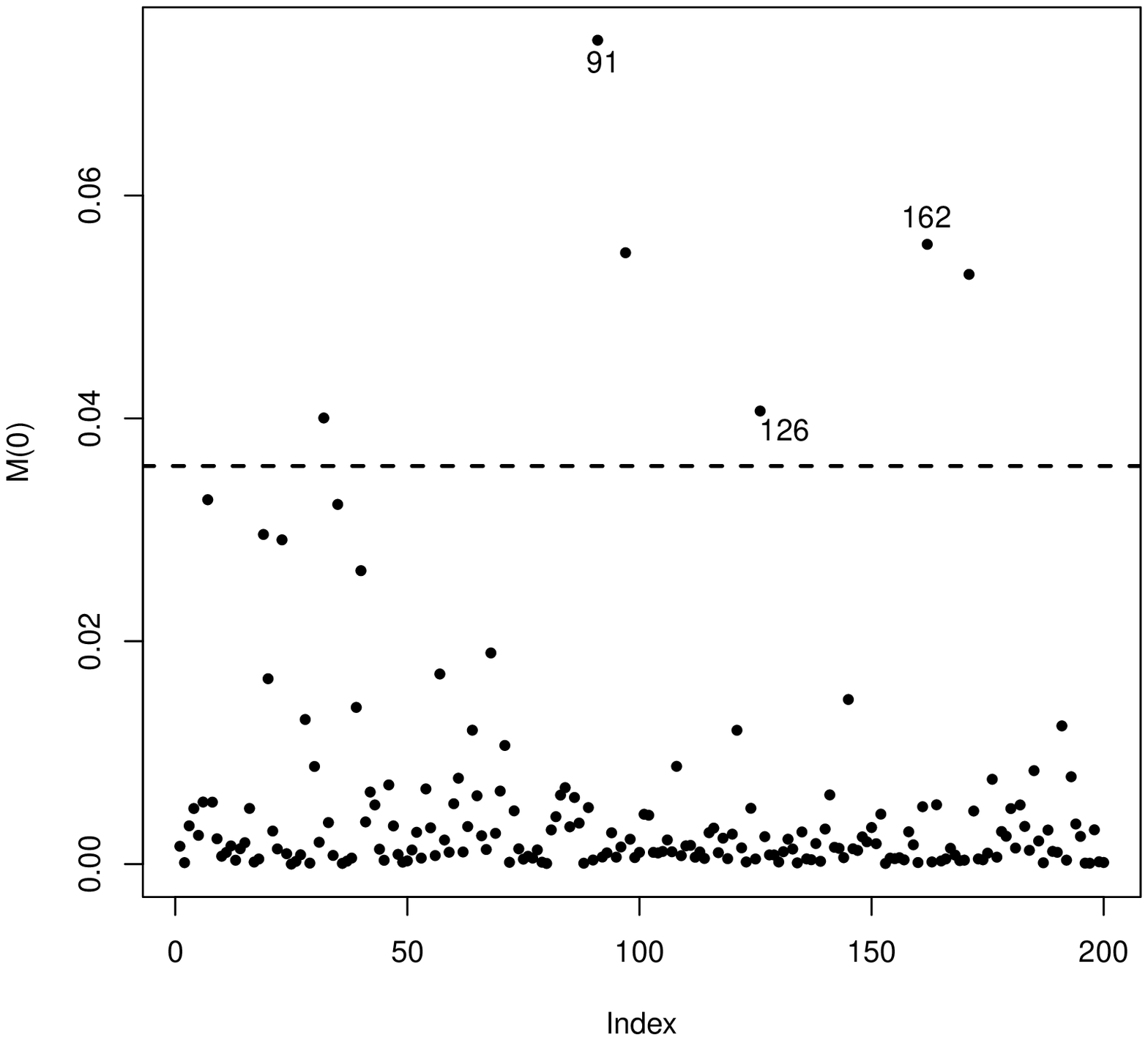}}
\subfigure[]{\includegraphics[width = 0.325\textwidth]{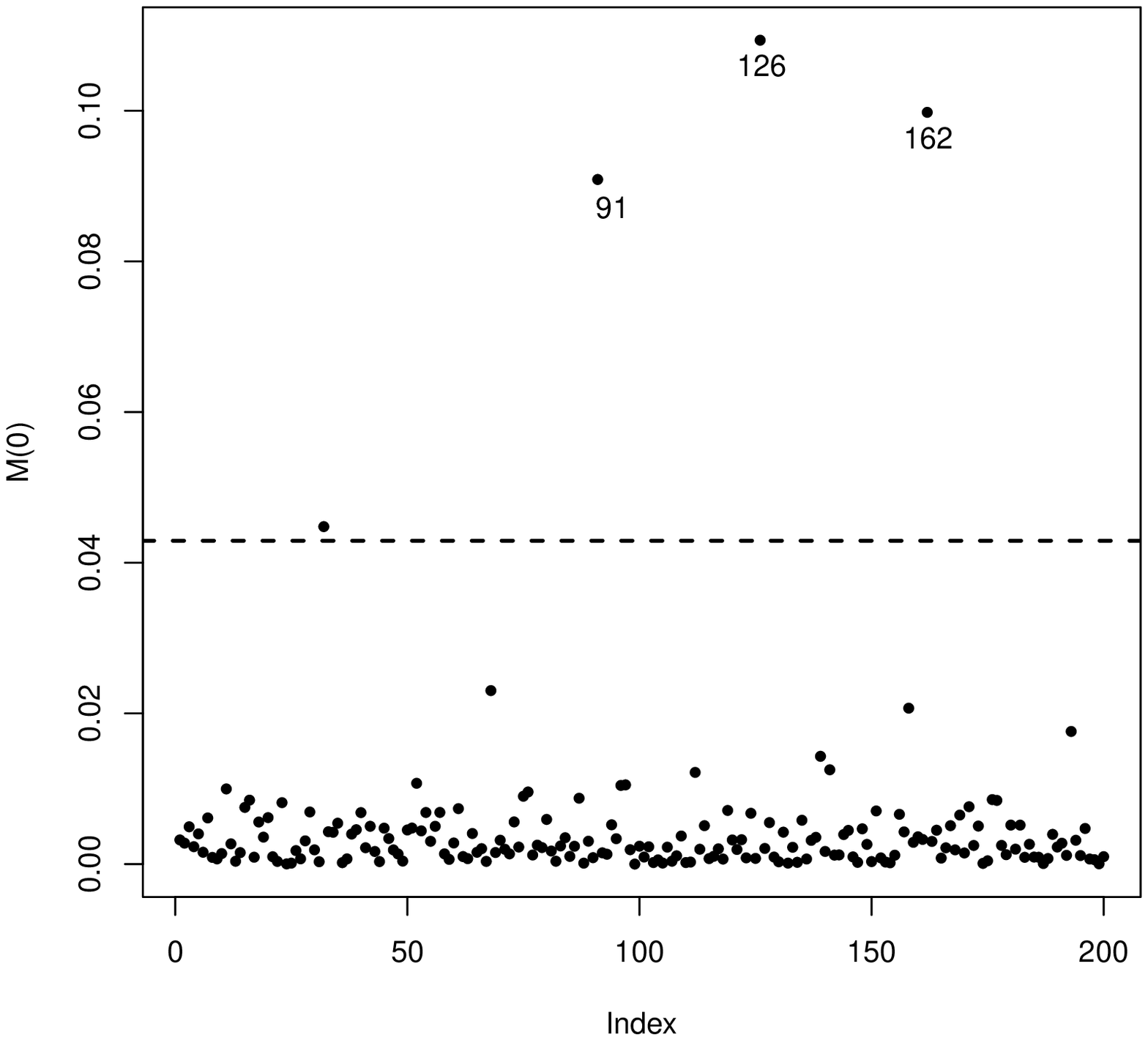}}

\caption{Local influence diagnostic plots for spherical (first row) and Mat\'{e}rn (second row) processes under (a,d) response perturbation (b,e) scale matrix perturbation and (c,f) explanatory variable perturbation schemes} \label{figu6}
\end{figure}

Figure \ref{figu6} (provided by the \texttt{localinfmeas} function) shows the diagnostic plots for the two simulated datasets. It can be seen that the artificial outliers were detected, as expected. For the response perturbation, the procedure detected two atypical observations under the spherical covariance structure and one under the Mat\'{e}rn covariance structure.  Thus, we can conclude that the proposed local influence measures work quite well to detect atypical observations in the context of censored spatial models.
%%%%%%%%%%%%%%%%%%%%%%%%%%%%%%%%%%%%%%%%%%%%%%%%%%%%%%%%%%%%%%%%%%%%%%%%%%%%%%%
%%%%%%%%%%%%%%%%%%%%%%%%%%%%%%%%%%%%%%%%%%%%%%%%%%%%%%%%%%%%%%%%%%%%%%%%%%%%%%%

\section{Conclusions}

This work introduces an \texttt{R} package for censored spatial data analysis that allows estimation, prediction and local influence diagnosis for censored Gaussian processes considering different linear trends and covariance structures. We  discuss in detail all functions available in this package as well as the theoretical framework. Plots provided by the package constitute one of the most powerful tools for comparing models, assessing predictions and analyzing local influence. Applications and several examples can be found in the manual of the \texttt{CensSpatial} package, available in the CRAN repository. To the best of our knowledge, this R package is the first proposal, available to practitioners, to perform statistical analysis of censored spatial data.

Future extensions of this work include the use of scale mixtures of normal (SMN) distributions to accommodate heavy-tailed features \citep{de2014influence}. We can also extend the proposal of \citet{diggle1998model} by including censoring in generalized linear mixed models (GLMMs).

%%%%%%%%%%%%%%%%%%%%%%%%%%%%%%%%%%%%%%%%%%%%%%%%%%%%%%%%%%%%%%%%%%%%%%%%%%%%%%%
%%%%%%%%%%%%%%%%%%%%%%%%%%%%%%%%%%%%%%%%%%%%%%%%%%%%%%%%%%%%%%%%%%%%%%%%%%%%%%%
\section*{Acknowledgments}
Christian E. Galarza acknowledges support from FAPESP-Brazil (Grant 2015/17110-9).\\

\bibliographystyle{chicago}
\bibliography{references}

\end{document}